\documentclass[12pt]{article} 

\usepackage{arxiv}

\usepackage[utf8]{inputenc} 
\usepackage[T1]{fontenc}    
\usepackage[hidelinks]{hyperref}       
\usepackage{url}           
\usepackage{booktabs}       
\usepackage{amsfonts}       
\usepackage{nicefrac}       
\usepackage{microtype}      
\usepackage{lipsum}		
\usepackage{graphicx}
\usepackage[super]{natbib}
\usepackage{doi}

\usepackage{siunitx}
\usepackage{adjustbox}
\usepackage{float}
\usepackage{caption}
\usepackage{amsmath}
\usepackage{booktabs}
\usepackage{lscape}
\usepackage{rotating}
\usepackage{setspace}

\usepackage{pdfpages}

\setcitestyle{comma}

\title{Target Aggregate Data Adjustment Method for Transportability Analysis Utilizing Summary-Level Data from the Target Population}

\author{ \href{https://orcid.org/0009-0006-6877-5293}{\includegraphics[scale=0.06]{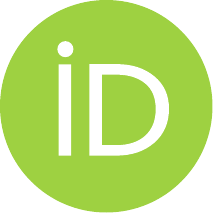}\hspace{1mm}Yichen Yan}
        \\
        Core Clinical Sciences, Vancouver, BC, Canada \\
	Department of Statistical and Actuarial Science, Simon Fraser University \\
        Burnaby, BC, Canada \\
	\texttt{yya276@sfu.ca} \\
	\And
    {\hspace{1mm}Quang Vuong} \\
	Core Clinical Sciences\\
        Vancouver, BC, Canada \\
	\texttt{quang@coreclinicalsciences.com} \\
	\And
    {\hspace{1mm}Rebecca K Metcalfe} \\
	Core Clinical Sciences, Vancouver, BC, Canada\\
        Centre for Advancing Health Outcomes, University of British Columbia \\
        Vancouver, BC, Canada \\
	\texttt{rebecca@coreclinicalsciences.com} \\
    	\And
    {\hspace{1mm}Tianyu Guan} \\
	Department of Mathematics and Statistics, York University \\
        North York, ON, Canada \\
	\texttt{tguan@yorku.ca} \\
	\And
	\href{https://orcid.org/0000-0002-3620-9717}
    {\includegraphics[scale=0.06]{orcid.pdf}\hspace{1mm}Haolun Shi} \\
	Department of Statistical and Actuarial Science, Simon Fraser University \\
        Burnaby, BC, Canada \\
	  \texttt{haolun\_shi@sfu.ca} \\
        \And
    {\hspace{1mm}Jay JH Park*} \\
	Core Clinical Sciences, Vancouver, BC, Canada \\
        Department of Health Research Methodology, Evidence, and Impact, McMaster University \\
        Hamilton, ON, Canada\\
	\texttt{parkj136@mcmaster.ca} \\
}

\renewcommand{\shorttitle}{\small TADA: Target Aggregate Data Adjustment Method for Transportability Analysis}

\hypersetup{
pdftitle={Target Aggregate Data Adjustment Method for Transportability Analysis},
pdfsubject={q-bio.NC, q-bio.QM},
pdfauthor={Yichen ~Yan, et al.},
pdfkeywords={transportability analysis; causal inference; survival analysis; aggregate-level data; inverse probability of censoring weights; method of moments},
}

\begin{document}
\maketitle
\newpage

\begin{abstract}
\setstretch{1.3}
    Transportability analysis is a causal inference framework used to evaluate the external validity of studies by transporting treatment effects from a study sample to an external target population by adjusting for differences in the distributions of their effect modifiers. Most existing methods require individual patient-level data (IPD) for both the source and the target population, narrowing its applicability when only target aggregate-level data (AgD) are available. For survival analysis, accounting for censoring may be needed to reduce bias, yet AgD-based transportability methods in the presence of informative-censoring remain underexplored. Here, we propose a two-stage weighting framework named ``Target Aggregate Data Adjustment'' (TADA) that can simultaneously adjust for both censoring bias and distributional imbalances of effect modifiers. In our framework the final weights are the product of the time-varying inverse probability of censoring weights and participation weights derived using the method of moments. We have conducted an extensive simulation study to evaluate TADA's performance. We have applied our methods to a real case study on the squamous non-small-cell lung cancer trial (NCT00981058). Our results indicate that TADA can effectively control the bias resulting from moderate censoring representative of most practical scenarios, and enhance the application and clinical interpretability of transportability analyses in settings with limited data availability.
\end{abstract}


\newpage
\section{Introduction}
\label{sec1}
\setstretch{1.3}
Transportability analysis is a novel causal inference framework used to quantitatively assess the external validity of randomized or observational studies~\citep{degtiar2023review,excellence2013guide,cole2010generalizing,stuart2011use,atkins2011assessing,wang2019using}. 
Transportability analysis methods ``transport'' findings from a source sample, gathered from a randomized clinical trial (RCT) or non-randomized (observational) study to an external target population by adjusting for differing distributions of effect modifiers between the source and target samples~\citep{excellence2013guide,cole2010generalizing,stuart2011use,atkins2011assessing,wang2019using}. 
By evaluating external validity, transportability analyses help bridge the gap between effect estimates observed in a source population and their anticipated impact in a target population of interest~\citep{williams2020external}. 

While different transportability methods are available, most require individual participant-level data (IPD) for both the source and target samples~\citep{cole2010generalizing,buchanan2018generalizing,tipton2013stratified,yang2020doubly,dahabreh2020extending}. Access to such data is often limited, presenting practical challenges to transportability analyses~\citep{degtiar2023review}. A common scenario is that IPD is only available for the source study, while the target population data is limited to aggregate-level data (AgD) available from publications. Recent methods developments have sought to leverage AgD to improve the feasibility of transportability analyses. For instance, entropy balancing approaches use convex optimization to derive sampling weights that align covariate moments between IPD from the source sample and AgD from the target population, facilitating effect estimate transport. This approach has been refined with calibration techniques, including an exponential‐tilting calibration scheme which achieves semiparametric efficiency under mild regularity conditions~\citep{josey2021transporting, josey2022calibration}. In another study by Quan et al.~\citep{quan2025generalizing}, they introduced a double inverse probability weighting (DIPW) method tailored for situations where only the summary data is available due to privacy constraints of specific countries/regions. Other proposed methods incorporate more flexible weighting strategies that use individual-level covariates from the source sample to seeks additional covariate balance between the treated and control groups~\citep{chen2023entropy}.

Existing studies on these AgD-based transportability methods have explored binary and continuous outcomes, and there has been limited exploration for time-to-event outcomes. Given the importance of accurately applying trial results to broader populations, recent work has explored different ways to transport survival outcomes from an RCT to target populations. For example, Ramagopalan et al.~\citep{ramagopalan2022transportability} developed a method for transporting survival estimates: they transported overall survival (OS) effect estimates from US clinical practice to the Canadian population using baseline covariates adjustments based on the Canadian population, but did so with individual-level data available in both populations. Zuo et al.~\citep{zuo2022transportability} applied proportional hazards and accelerated failure time models to transport survival outcomes in adjuvant colon cancer trials. They explored one-trial-at-a-time and leave-one-trial-out methods, demonstrating how varying model structures and moment adjustments can address treatment effect discrepancies due to heterogeneous covariate distributions across trials, providing a structured framework for extending survival results to diverse populations. However, the framework likewise presumes the availability of individual-level data for both the source and each target trial. Lee et al.~\citep{lee2024transporting} extended the augmented calibration weighting (ACW) methods with the linear spline-based hazard regression model to generalize treatment effects from the ACTG 175 HIV trial to various target populations without the limitation of proportional hazard (PH) assumption. This extension requires target-population IPD (or an equivalent synthetic micro-dataset), imposing the same practical constraint as mentioned.

Informative censoring in survival analysis can critically impact the performance of statistical methods. For transportability analyses, this is no different. There has been recent work developing methods to account for censoring when performing transportability analysis on survival data. One such work is by Berkowitz et al.~\citep{berkowitz2018generalizing}, which uses inverse odds of selection weights with survival tools like the Kaplan–Meier estimator and marginal structural Cox models to transport survival outcomes. Although effective for estimating counterfactual survival times, this method assumes censoring is independent of covariates, potentially limiting its real-world applicability. Lee et al.~\citep{lee2022doubly} advance this by addressing covariate-dependent censoring in survival transportability. Using a doubly robust (DR) estimator, their method integrates inverse probability weighting for sampling, censoring, and treatment assignment with outcome regression, ensuring accuracy even when censoring depends on covariates. Cao et al.~\citep{cao2024transporting} expanded on weighting and DR methods to estimate counterfactual survival functions within the target population. Unlike Lee et al.'s calibration weighting estimators, this approach employs simpler estimators that rely on direct weight estimation or survival outcome modelling. These methods have advanced survival outcome transportability, but they still require access to IPD of the target population. There is a current gap in transportability analysis methods that can account for covariate-dependent censoring in settings where only AgD is available in the target population. 

In this paper, we propose a novel framework called ``Target Aggregate Data Adjustment'' (TADA) that can account for covariate-dependent time-varying censoring when transporting findings to target AgD. Our TADA framework can incorporate inverse probability of censoring weights with the method of moments (MoM) weighting to account for differences in effect modifier distributions between source and target populations. TADA then derives final matching weights for each individual in the source dataset through a two-stage scheme, balancing the source and target populations to transport treatment effects. The R codes to implement TADA and reproduce the simulation results can be obtained from https://github.com/CoreClinicalSciences/TADA. TADA is also available as a main module in \texttt{TransportHealth}, a recently published open-source R package developed by Park et al.~\citep{park2025introducing} for transportability analyses and causal reasoning.

The paper is organized as follows. In Section~\ref{sec2}, we introduce the notations and formalize the causal inference framework for survival outcomes in transportability with necessary assumptions. Section~\ref{sec3} introduces the methodological approaches used in TADA. Sections~\ref{sec4} and~\ref{sec5} cover the simulation settings and present the results, respectively. Sensitivity analyses are provided in Section~\ref{sec6}. In Section~\ref{sec7}, we apply TADA to transport the Kaplan–Meier survival curve of a control arm from the stage IV squamous non-small-cell lung cancer (NSCLC) treatment with the inhibitor of EGF receptor (SQUIRE) trial (NCT00981058)~\citep{thatcher2015necitumumab} to a real-world target population with stage IV NSCLC in Ontario, Canada~\citep{seung2020real}. A discussion is provided in Section~\ref{sec8}, with concluding remarks in Section~\ref{sec9}. In the Supplementary Materials, we provide a brief review of existing transportability methods for aggregate target data, a theoretical justification of unbiased estimation of TADA and additional simulation results.

\section{Notation, Framework and Assumptions}
\label{sec2}
\setstretch{1.3}
Suppose we are interested in comparing the effectiveness of two treatments. We have access to the source sample $\mathcal{S}$ of size $N_{\textit{s}}$, with individual data. Let $\textbf{X} \in \mathcal{X}$ be a vector of baseline covariates, consisting of two subsets: (1) covariates $\mathbf{L}$, which include variables associated with the risk of being censored (e.g., dropout or loss to follow-up) and required to ensure conditional exchangeability of counterfactual survival times and censoring; and (2) effect modifiers $\mathbf{E}$, which include variables that influence how treatment effects may vary across different populations and required to ensure conditional exchangeability of counterfactual survival times and trial participation. For simplicity and clarity in presentation, we uniformly denote both subsets together as $\mathbf{X}$ throughout the manuscript. Let $A \in \{0, 1\}$ be the treatment indicator, with $A = 1$ referring to active treatment and $A = 0$ representing control. Let $S \in \{0,1\}$ be the trial participation indicator, with $S=1$ referring to trial participation in the RCT, and $S=0$ otherwise. Let $T$ denote the survival time, defined as the event occurrence time. Following the potential outcomes framework~\citep{rubin1974estimating,rubin1986comment}, let $T^{(a)}$ be the potential survival time if a subject receives the treatment $A = a$. Under the stable unit treatment value assumption (SUTVA)~\citep{rubin1974estimating}, the survival time is indicated as $T = T^{(1)}A + T^{(0)}(1-A)$. Let $C$ be the censoring time. In the presence of right censoring, the survival time $T_i$ is not always observed for all subjects; hence, we define observed time as $U_i = T_i \wedge C_i$ and censoring indicator $\Delta_i = I(T_i \leq C_i)$, where $\wedge$ represents the minimum of two values, and $I(\cdot)$ is an indicator function. In summary, from the RCT source sample, we observe $\{U_i, \Delta_i, A_i, \textbf{X}_i, S_i = 1\}$ from $i = 1,\cdots, N_\textit{s}$ subjects. 

The target population, $\mathcal{T}$, contains $N_\textit{t}$ individuals. We do not observe either the treatment or outcome but have access to the sample moments $\overline{h}_{k, \mathcal{T}} = \frac{1}{N_\textit{t}} \sum_{i \in \mathcal{T}} h_k(\mathbf{X}_i)$ of certain covariate functions $h_k: \mathcal{X} \rightarrow \mathbb{R} $, where $k = 1, \dots, K$. 

Let $S_a(t) = \Pr(T^{(a)} \geq t)$ be the survival function and $\lambda_a(t) = \text{lim}_{h \rightarrow 0} \frac{1}{h}\frac{\Pr(t \leq T^{(a)} \leq t+h)}{\Pr(T^{(a)} \geq t)}$ be the hazard function. Under the proportional hazards assumption, i.e., $\lambda_1(t) / \lambda_0(t)$ being a constant, the estimand of interest is chosen as the marginal hazard ratio, derived by estimating the causal effect of treatment on survival time within a weighted Cox PH model. 

In RCTs, treatment assignment $A$ is independent of covariates \(\textbf{X}\) given study participation $S$, thus the treatment propensity score \(\pi_{A}\) is defined as the fixed probability of receiving the treatment. Let $\pi_{S}(\textbf{X}) = \Pr(S = 1 \mid \textbf{X})$ be the sampling score. We make the following assumptions for the identification of the estimand of interest:
\\

\textbf{Assumption 1.} (Randomization) $\{T^{(0)},T^{(1)}\} \perp\!\!\!\perp A \mid S = 1$; and $0 < \pi_{A} < 1$.

\textbf{Assumption 2.} (Ignorability and positivity of trial participation) $\{T^{(0)},T^{(1)}\} \perp\!\!\!\perp S \mid \textbf{X}$; and $0 < \pi_{S}(\textbf{X}) < 1$ for all values of $\textbf{X}$.

\textbf{Assumption 3.} (Noninformative censoring conditional on covariates and treatment) $\{T^{(0)},T^{(1)}\} \perp\!\!\!\perp C \mid (\textbf{X},A,S = 1)$, which also implies $T \perp\!\!\!\perp C \mid (\textbf{X},A,S = 1)$.  
\\

Given the inherent difficulties in directly validating these assumptions, relevant domain knowledge should support their practical plausibility. Assumption 1 holds in the RCT by design. If all information related to trial participation and outcomes is captured in the data, then the first part of Assumption 2 is reasonable. Additionally, the second part of Assumption 2 requires that all subgroups defined by covariates have a non-zero probability of trial participation. Specifically, if certain patient characteristics are completely unobserved or excluded by design, then individuals with these characteristics would have no possibility of participating in the trial, violating the positivity condition and limiting the transportability of trial results to such populations~\citep{lee2023improving}. Assumption 3 is a common assumption in survival analysis and imposes fewer restrictions compared to the conditional independence assumption between censoring and survival time based solely on treatment~\citep{buchanan2018generalizing,berkowitz2018generalizing,hernan2010hazards,ackerman2021generalizing}.

\section{Target Aggregate Data Adjustment (TADA)}
\label{sec3}
\setstretch{1.3}

\subsection{Participation Weights: Method of Moments}

When only target AgD is available, conducting transportability analysis faces challenges due to limited information that is provided by AgD. To address this issue, we propose a MoM-based approach as a feasible alternative for covariate adjustment. 

The original application of MoM methods was in matching-adjusted indirect comparison (MAIC) to achieve balance of covariates between two RCTs of comparison~\citep{signorovitch2010comparative}. For each covariate of interest, the MoM approach aligns the sample moments calculated from the IPD of the source population with the corresponding moments from the AgD of the target population. Greater availability of AgD, particularly higher-order target sample moments, reduces bias in the population matching process. For instance, the access to additional summary statistics such as the sample standard deviation, median, proportion, or even higher-order moments (e.g., skewness or kurtosis) for covariates can result in more accurate estimators compared to relying solely on the sample mean~\citep{chen2023entropy}. Our approach is conceptually similar to the strategy introduced by Josey et al.~\citep{josey2021transporting} where entropy balancing was employed to achieve the same goal. The existing literature highlights the connection between entropy balancing and the MoM approach, particularly in applications to ITC and transportability analyses. Both methods are statistically equivalent, sharing common advantages and robustness~\citep{phillippo2020equivalence}, while MoM has the advantage of being relatively intuitive in clinical interpretation, as discussed in section S1 of Supplementary Materials.

Here, we illustrate the theoretical process of participation weight estimation by the MoM approach through matching first-order sample moments (sample means). For simplicity and ease of illustration, we describe the process for a single covariate component, while the approach naturally extends to multiple covariates. Consider a specific covariate component $X$ available in both the source and target populations. Assume that we have IPD for $X$ in the source sample, but only the sample mean $\overline{X}$ for this covariate in the target sample. 

For each individual $i$, let $X_{i}$ represent the individual-level covariate data available in the source sample. We construct participation weights in the widely used exponential tilting form~\citep{josey2021transporting,chen2023entropy, hainmueller2012entropy, signorovitch2012matching}. Formally, we model participation weights for individual \(i\) as:
    \[
    w_i^{\text{MoM}} = \exp\bigl(\gamma + X_i\cdot\zeta\bigr),
    \]
where $\gamma$ is a normalizing constant and $\zeta$ is a parameter to be determined such that the weighted source sample mean matches the target sample mean. Then MoM-based method matches the weighted source sample mean with the target sample mean:
\[
\frac{\sum_{i=1}^{N_\textit{s}}X_i\cdot w_i^{\text{MoM}}}{\sum_{i=1}^{N_\textit{s}}w_i^{\text{MoM}}} = \overline{X}.
\]

Without loss of generality, we apply a transformation by centering the covariate \(X\) in the source population, using the target population mean \(\overline{X}\). This yields a new covariate \(X_{i,\text{CENTERED}} = X_i - \overline{X}\), resulting in the simplified condition where \(\overline{X}_{\text{CENTERED}} = 0\) in the target population and naturally cancels out the intercept term $\gamma$ without separate estimation. Thus, the weights simplify to:
\[
    w_i^{\text{MoM}} = \exp\bigl(X_{i,\text{CENTERED}}\cdot\zeta\bigr),
\]
and the equality constraint becomes:
\[
    \frac{\sum_{i=1}^{N_\textit{s}}X_{i,\text{CENTERED}}\cdot w_i^{\text{MoM}}}{\sum_{i=1}^{N_\textit{s}}w_i^{\text{MoM}}} = 0.
\]

By further simplifying the constraint as $\sum_{i=1}^{N_\textit{s}}X_{i,\text{CENTERED}}\cdot w_i^{\text{MoM}} = 0 = Q'(\zeta)$, where the objective function \(Q(\zeta)\) is given by $Q(\zeta) = \sum_{i=1}^{N_\textit{s}}\exp(X_{i,\text{CENTERED}}\cdot\zeta)$, the solution for \(\zeta\) that satisfies this equality constraint is unique and globally optimal due to the convexity of \(Q(\zeta)\). Thus, stable estimates of participation weights are obtained.

\subsection{Censoring Weights: Inverse Probability of Censoring Weight} 
In longitudinal studies, participants are lost to follow-up for various reasons that may be related to the treatment or competing events. This loss of follow-up can result in systematic differences between those who remain in the study and those who do not across treatment groups. While RCTs generally eliminate confounding through randomization, censoring can still introduce bias if the probability of being censored is related to the outcome the participant would have experienced without censoring. To address this issue, censoring weights are applied during the estimation process, aiming to recreate a hypothetical population that would have been observed without censoring. TADA applies inverse probability censoring weights (IPCW) to account for bias resulting from censoring for the time-to-event outcome in the source population.

For each individual, the IPCW is time-varying. Specifically, at each observed event or censoring time point $t$, we estimate:
\[
  w_i^c(t) \;=\; \frac{1}{\Pr\bigl(\Delta_i(t)=1 \mid \mathbf X_i, A_i, T_i \ge t\bigr)},
\]
where \(\Delta_i(t)=1\) denotes not censored for individual \(i\) at time \(t\). These weights are calculated at each distinct event time during follow-up.

We use the Cox PH model~\citep{cox1972regression, therneau2000cox} to model the hazard of censoring conditional on covariates:
\begin{equation} 
    h_c(t \mid A_i, \textbf{X}_i) = h_{c0}(t) \cdot \exp\bigl\{\eta_i(t)\bigr\}, 
    \nonumber
\end{equation}
where \(h_c(t \mid A_i,\textbf{X}_i)\) is the hazard function for censoring at time \(t\) for individual \(i\), \(h_{c0}(t)\) is the time-varying baseline hazard function, and $\eta_i(t) = \beta_c^\top (A_i,\textbf{X}_i)$ is the linear predictor where \(\beta_c\) is the vector of regression coefficients and \((A_i,\textbf{X}_i)\) is the vector of treatment and covariates.

We estimate the baseline cumulative hazard function \(H_\text{c0}(t)\) using the Breslow estimator~\citep{breslow1972discussion}:
\begin{equation}
    H_\text{c0}(t) = \sum_{t_j \leq t} \frac{d_j}{\sum_{l \in R(t_j)} \exp\bigl\{\eta_i(t)\bigr\}},
    \nonumber
\end{equation}
where \(d_j\) is the number of censoring events at time \(t_j\), and \(R(t_j)\) is the risk set at time \(t_j\).

The censoring weight \(w_i^c(t)\) for individual \(i\) at time \(t\) is defined as the inverse of their survival probability at that time:
\begin{equation} 
    w^\textbf{c}_i(t) = \frac{1}{S_c\bigl(t \mid A_i,\textbf{X}_i\bigr)},
    \nonumber
\end{equation}
where $S_c\bigl(t \mid A_i,\textbf{X}_i\bigr) = \exp\bigl\{ -H_{c0}(t) \cdot \exp\bigl(\eta_i(t)\bigr)\bigr\}$ represents the uncensoring probability at time \(t\) given covariates and treatment. 

\subsection{Final Weights: A Two-staged Scheme}
We introduce the final weight for each individual \(i\) at time $t$ as the product of the participation weight \(w_i^{\mathrm{MoM}}(t)\) and the censoring weight \(w_i^{\mathbf{c}}(t)\), ensuring both population matching and bias correction for censoring as:
\[
  w_i^{\textbf{final}}(t)
  \;=\;
  w_i^{\textbf{MoM}}(t)
  \times
  w_i^{\textbf{c}}(t).
\]

When the distribution of the weights is highly skewed, it may be appropriate to consider trimming weights above a certain percentile (e.g., the $95^\text{th}$ or $99^\text{th}$). Such decisions should be guided by diagnostic assessments, including the inspection of weight distribution and sensitivity analyses across different truncation thresholds, to balance variance reduction against the risk of introducing bias. In the following simulation study introduced in Section~\ref{sec4}, we apply an overall $95^\text{th}$ percentile trimming threshold after checking the distribution histogram of raw final weights in various settings to achieve the optimal performance for most of scenarios. A systematic sensitivity analysis about various final weights truncation strategies is also conducted and results are presented in Section~\ref{sec6}.

This combined weighting approach ensures unbiased treatment effect estimates concerning both censoring and population differences. The weighted source population should closely resemble the target population, demonstrating population balance and supporting the treatment effect estimates. We provide a theoretical justification of unbiased estimation of TADA in section S2 of the Supplementary Materials.

In TADA, we employ a marginal structural model (MSM) to estimate the causal effect of treatment on survival outcomes. Since the weights used in TADA depend solely on baseline covariates and do not involve any time-varying components, we adopt a simplified MSM suitable for this context~\citep{robins2000marginal, hernan2000marginal}, while TADA remains agnostic to the specific outcome model or assumptions. The MSM is defined as a marginal Cox PH model~\citep{cox1972regression}:
\begin{equation} 
    \lambda_a(t \mid A_i) = h_0(t) \exp(\beta A_i), 
    \nonumber
\end{equation}
where $\lambda_a(t \mid A_i)$ is the hazard function at time $t$ for individual $i$ with treatment assignment $A_i$, $h_0(t)$ is the baseline hazard function, and $\beta$ represents the log hazard ratio of the treatment effect.

We estimate $\beta$ using a weighted partial likelihood approach, with the final weights $w_i^{\textbf{final}}(t)$ obtained from the two-stage scheme as detailed in Sections 3.1 to 3.3. Applying these final weights creates a pseudo-population in which: (i) the participation weights balance effect modifiers between the source and target populations, ensuring transportability of treatment effects; (ii) the censoring weights reduce bias from incomplete follow-up by assigning greater weights to individuals with higher risk of dropout, approximating the outcomes that would have been observed without censoring under Assumption 3; (iii) treatment is independent of baseline covariates. This framework enables us to estimate the transported treatment effect on the target population using only aggregate-level data.

The weighted partial likelihood function for the Cox model~\citep{binder1992fitting} is given by:
\begin{equation} 
    L(\beta) = \prod_{i=1}^{N_\textit{s}} \left[ \frac{\exp(\beta A_i)}{\sum_{j \in R(t_i)} w_j^\textbf{final}(t_i) \exp(\beta A_j)} \right]^{\Delta_i w_i^\textbf{final}(t_i)}, 
    \nonumber
\end{equation}
where $N_\textit{s}$ is the number of individuals in the source sample, $\Delta_i$ is the censoring indicator for individual $i$, and $R(t_i)$ is the risk set of individual $i$ at time $t$. Maximizing this weighted likelihood function provides an estimate of $\beta$, the log hazard ratio, which quantifies the causal effect of the treatment in the target population. We are estimating the variance of $\beta$ using a nonparametric bootstrap method.

\section{Simulation Study}
\label{sec4}
\setstretch{1.3}

The simulation study applies the ADEMP framework proposed by Morris et al.~\citep{morris2019using} to comprehensively evaluate the effectiveness of TADA in transportability analysis using AgD for a target population, specifically examining its performance in adjusting for censoring. Our primary goal (\textbf{A}im) is to assess the estimated causal effects of treatment on overall survival across different RCT scenarios, with and without censoring adjustments made via TADA. Detailed descriptions of the \textbf{D}ata-generating mechanisms, \textbf{E}stimands, \textbf{M}ethods, and \textbf{P}erformance measures are provided in the following subsections.

\subsection{Data-Generating Mechanism}
In each simulation replicate, we generate three datasets named \texttt{studyIPD}, \texttt{targetAgD}, and \texttt{targetPseudoIPD} respectively: \texttt{studyIPD} contains simulated individual-level data from the source population, including treatment status, baseline covariates, event times, censoring times, observed times, and indicators for observed and censored events. \texttt{targetAgD} provides aggregate summary statistics (such as proportions, means, and standard deviations of each baseline covariate) for the target population, matching those in \texttt{studyData}. \texttt{targetPseudoIPD} consists of pseudo individual-level data for the target population, including baseline covariates, treatment status, and event times. We generate both event and censoring times from Weibull distributions, implying that the baseline hazard functions are time-varying~\citep{kalbfleisch2011statistical, collett2023modelling}. Specifically, the Weibull hazard function is defined as:
\begin{equation} 
    h(t) = \lambda \alpha t^{\alpha - 1}, 
    \nonumber
\end{equation}
where $\lambda$ is the scale parameter, $\alpha$ is the shape parameter, and $t$ represents time. If $\alpha > 1$, the hazard increases over time; if $\alpha < 1$, the hazard decreases. Correspondingly, the cumulative hazard and survival functions are:
\begin{equation} 
    H(t) = \lambda t^\alpha, \quad S(t) = \exp\left\{-H(t)\right\} = \exp\left(-\lambda t^\alpha\right). 
    \nonumber
\end{equation}

To generate \texttt{studyIPD}, we first simulate a source population of 50,000 individuals and then randomly sample 200 individuals for the final dataset. Each individual has three baseline covariates: $X_1$ and $X_2$ are binary variables generated from binomial distributions with probabilities 0.45 and 0.65, respectively. $X_3$ is a continuous variable generated from a standard normal distribution (mean 0, variance 1). Treatment status is assigned randomly to simulate an RCT scenario, with an allocation rate of 1:1.

Event times are generated from a Weibull distribution, incorporating the effects of covariates, treatment and the interaction between treatment and each covariate through the linear predictor $\eta_{\text{event}, i}$.
\begin{equation} 
    \begin{aligned} 
    \eta_{\text{event}, i} &= \beta_{X_1} X_{1,i} + \beta_{X_2}  X_{2,i} + \beta_{X_3} X_{3,i} 
    + \beta_{\text{trt}} \cdot \text{Treatment}_i \\ &+ \beta_{\text{trt-$X_1$}} \cdot \text{Treatment}_i \cdot X_{1,i} + \beta_{\text{trt-$X_2$}} \cdot \text{Treatment}_i \cdot X_{2,i} + \beta_{\text{trt-$X_3$}} \cdot \text{Treatment}_i \cdot X_{3,i}, 
    \nonumber
    \end{aligned} 
\end{equation}
where the main effects are set as $\beta_{X_1} = 0.5$, $\beta_{X_2} = -0.3$, $\beta_{X_3} = 0.2$, and $\beta_{\text{trt}} = -0.4$.

We generate a total of 10 distinct random scenarios by varying the coefficient combinations of the treatment-covariate interaction terms $(\beta_{\text{trt-$X_1$}}, \beta_{\text{trt-$X_2$}}, \beta_{\text{trt-$X_3$}})$, while keeping the main effect coefficients $\beta_{X_1}$, $\beta_{X_2}$, $\beta_{X_3}$ and $\beta_{\text{trt}}$ constant. This approach ensures that all variables are sufficiently perturbed during data generation, creating a range of representative and complex simulation scenarios. The coefficient combinations are provided in Table~\ref{table1} by scenarios.
\begin{table}[h!]
\renewcommand{\arraystretch}{1.4}
    \centering
    \small
    \begin{tabular}{cccccc}
    \toprule
    Scenario & $\beta_{\text{trt-$X_1$}}$ & $\beta_{\text{trt-$X_2$}}$ & $\beta_{\text{trt-$X_3$}}$ \\
    \midrule
    1 & 1.20 & 0.35 & 1.30 \\
    2 & 0.58 & 0.35 & 1.30 \\
    3 & -0.90 & 0.35 & 1.30 \\
    4 & -0.60 & 0.35 & 1.30 \\
    5 & 1.20 & 0.20 & 1.30 \\
    6 & 1.20 & -0.70 & 1.30 \\
    7 & 1.20 & -0.10 & 1.30 \\
    8 & 1.20 & 0.35 & 0.45 \\
    9 & 1.20 & 0.35 & -0.30 \\
    10 & 1.20 & 0.35 & -0.85 \\
    \bottomrule
    
    \end{tabular}
    \caption{10 distinct random scenarios and corresponding coefficient combinations of the treatment-covariate interaction terms $(\beta_{\text{trt-$X_1$}}, \beta_{\text{trt-$X_2$}}, \beta_{\text{trt-$X_3$}})$.}
    \label{table1}
\end{table}

Event times $T_{i}$ are simulated using a Cox PH model with a baseline hazards function following a Weibull distribution as:
\begin{equation} 
    T_{i} = \left( -\frac{\ln(u_i)}{\lambda_{\text{event}} \exp\left( \eta_{\text{event}, i} \right)} \right)^{1/\alpha_{\text{event}}}, 
    \nonumber
\end{equation}
where shape parameter $\alpha_{\text{event}} = 1.5$ and scale parameter $\lambda_{\text{event}} = 0.1$. These parameters are chosen to reflect an increasing hazard over time ($\alpha_{\text{event}} > 1$). $u_i$ is a random variable drawn from the uniform distribution on (0, 1). 

In \texttt{studyIPD}, we generate censoring times for each individual similarly to event times. A linear predictor $\eta_{\text{censor}, i}$ is constructed for each individual to incorporate the effects of covariates on the hazard function.
\begin{equation} 
    \eta_{\text{censor}, i} = \beta_{0} + \beta_{X_1}^c X_{1,i} + \beta_{X_2}^c X_{2,i} + \beta_{X_3}^c X_{3,i} + \beta_{\text{trt}}^c \cdot \text{Treatment}_i , 
    \nonumber
\end{equation}
where $\beta_{X_1}^c = -0.2$, $\beta_{X_2}^c = 0.4$ and $\beta_{X_3}^c = -0.1$ as the main effect of baseline covariates, $\beta_{\text{trt}}^c = 0.25$ as the treatment main effect and $\beta_{0}$ as the intercept. We choose various candidates of $\beta_{0}$ to directly adjust the overall censoring proportion among \texttt{studyIPD}. We set $\beta_{0}$ as 2.5, 3.3, 3.7 and 4.3 respectively to obtain \texttt{studyIPD} with around 20\%, 30\%, 40\% and 50\% overall censoring proportion while keeping other coefficients being consistent.

The censoring times $C_{i}$ are generated using:
\begin{equation} 
    C_{i} = \left( -\frac{\ln(u_i')}{\lambda_{\text{censor}} \exp\left( \eta_{\text{censor}, i} \right)} \right)^{1/\alpha_{\text{censor}}}, 
    \nonumber
\end{equation}
where shape parameter $\alpha_{\text{censor}} = 1.5$ and scale parameter $\lambda_{\text{censor}} = 0.001$. $u_i'$ is an independent random variable from the uniform distribution on (0, 1).

For each individual, the observed time $U_{i}$ is determined by:
\begin{equation} 
    U_{i} = T_{i} \wedge C_{i},
    \nonumber
\end{equation}
where $\wedge$ represents the minimum of two values.

The censoring indicator $\Delta_i$ is defined as:
\begin{equation} 
    \Delta_i = 
    \begin{cases} 
        1, & \text{if } T_{i} \leq C_{i} \quad (\text{event observed}), \\
        0, & \text{if } T_{i} > C_{i} \quad (\text{censored}).
    \end{cases}
    \nonumber
\end{equation}

Similarly, we generate \texttt{targetAgD} by simulating a larger target population of 200,000 individuals and then randomly sampling 1,000 individuals for the final dataset. Each individual has the same three baseline covariates as in the source population ($X_1$, $X_2$, and $X_3$) with different distribution characteristics: $X_1$ is binary and generated from a binomial distribution with probability 0.35. $X_2$ is binary and generated from a binomial distribution with probability 0.55. $X_3$ is continuous and generated from a normal distribution with a mean of 0.21 and a variance of 1.5. Since \texttt{targetAgD} is an aggregate-level dataset, we summarize the proportions of $X_1$ and $X_2$, and the mean and standard deviation of $X_3$ as the final data.

The individual-level covariate data for \texttt{targetPseudoIPD} are generated concurrently before aggregation. To obtain the ``true'' treatment effect in the target population for benchmarking, we assign treatment status and generate event times for each individual in \texttt{targetPseudoIPD}. Treatment status is randomly assigned with a 50\% probability to simulate an RCT scenario. We use the same Weibull distribution parameters ($\alpha_{\text{event}}$, $\lambda_{\text{event}}$), the same linear predictor structure $\eta_{\text{event}, i}$ and coefficients as in \texttt{studyIPD}, applying them to \texttt{targetPseudoIPD} to ensure consistency.

\subsection{Estimands, Targets, and Methods}
We are interested in estimating the marginal hazard ratio in the target population. We consider two analytical strategies for handling censoring in the context of transportability analysis.
\begin{itemize}
    \item Adjusting censoring: In this case, TADA balances the source and target population with the participation weights introduced in Section 3.1, as well as adjusting the bias resulting from censoring via IPCW introduced in Section 3.2. The final weights are the product of the two as Section 3.3.
    \item Ignoring censoring: In this case, TADA ignores the impact of censoring and only balances the source and target population with the participation weights introduced in Section 3.1. The final weights are equivalent to the participation weights estimated by the method of moments.
\end{itemize}

For both strategies, we estimate the average marginal hazard ratios using a Cox PH model based on multiple replicates. We apply the Cox PH model on \texttt{targetPseudoIPD} to estimate average marginal hazard ratios as the pseudo ``true HR'' of the target population as the comparison benchmark.

\subsection{Performance Measures}
The performance measures of interest are the bias of the estimators calculated using the above methods and their coverage of 95\% confidence intervals compared to the analysis in the IPD target data. Let $\hat{\gamma}_{\text{cen,r}}$ be the estimated marginal hazard ratios for the $r = 1,2,\cdots, R$-th simulated replicate when considering the adjustment of censoring. Let $\hat{\gamma}_{\text{non-cen,r}}$ be the estimated marginal hazard ratios for the $r = 1,2,\cdots, R$-th simulated replicate when ignoring the adjustment of censoring. With the pseudo ``true HR'' of target population being $\overline{\gamma}$, the bias of two cases aforementioned are estimated with
\begin{equation}
    \widehat{\text{Bias}}_{\text{cen}} = \frac{1}{R}\sum_{r = 1}^R \left( \hat{\gamma}_{\text{cen,r}} - \overline{\gamma} \right),\quad
    \widehat{\text{Bias}}_{\text{non-cen}} = \frac{1}{R}\sum_{r = 1}^R \left(\hat{\gamma}_{\text{non-cen,r}} - \overline{\gamma}\right).
    \nonumber
\end{equation}

For each replicate \(r = 1, \ldots, R\), we construct a 95\% percentile‐based bootstrap confidence interval $\bigl[\widehat{\mathrm{CI}}_{\mathrm{low},r},\;\widehat{\mathrm{CI}}_{\mathrm{high},r}\bigr]$ for the hazard ratio, where
\begin{equation}
\widehat{\mathrm{CI}}_{\mathrm{low},r}
  = \text{Quantile}_{0.025}\bigl(\hat{\gamma}_{r}^{*(1)},\dots,\hat{\gamma}_{r}^{*(B)}\bigr),
\quad
\widehat{\mathrm{CI}}_{\mathrm{high},r}
  = \text{Quantile}_{0.975}\bigl(\hat{\gamma}_{r}^{*(1)},\dots,\hat{\gamma}_{r}^{*(B)}\bigr) \nonumber.
\end{equation}

We then define the coverage as the proportion of replicates for which $\overline{\gamma}$ lies within this interval:
\begin{equation}
    \widehat{\mathrm{Coverage}}_{\mathrm{cen}}
  = \frac{1}{R}\sum_{r=1}^{R}
      I\Bigl\{\overline{\gamma} \in \bigl[\widehat{\mathrm{CI}}_{\mathrm{low},r}^{(\mathrm{cen})},
    \widehat{\mathrm{CI}}_{\mathrm{high},r}^{(\mathrm{cen})}\bigr]
        \Bigr\}, \quad \widehat{\mathrm{Coverage}}_{\mathrm{non\text{-}cen}}
  = \frac{1}{R}\sum_{r=1}^{R}
      I\Bigl\{\overline{\gamma} \in \bigl[\widehat{\mathrm{CI}}_{\mathrm{low}, r}^{(\mathrm{non\text{-}cen})}, \widehat{\mathrm{CI}}_{\mathrm{high}, r}^{(\mathrm{non\text{-}cen})}\bigr] \Bigr\} \nonumber.
\end{equation}

Considering the trade-off between accuracy and computational cost in the presence of time-varying weight estimation in our simulation, we simulate $R$ = 500 replicates for each scenario and within each replicate generate $B = 200$ bootstrap resamples.

\section{Results}
\label{sec5}
\setstretch{1.3}

\begin{figure}
    \centering
    \includegraphics[width = 1.0\textwidth]{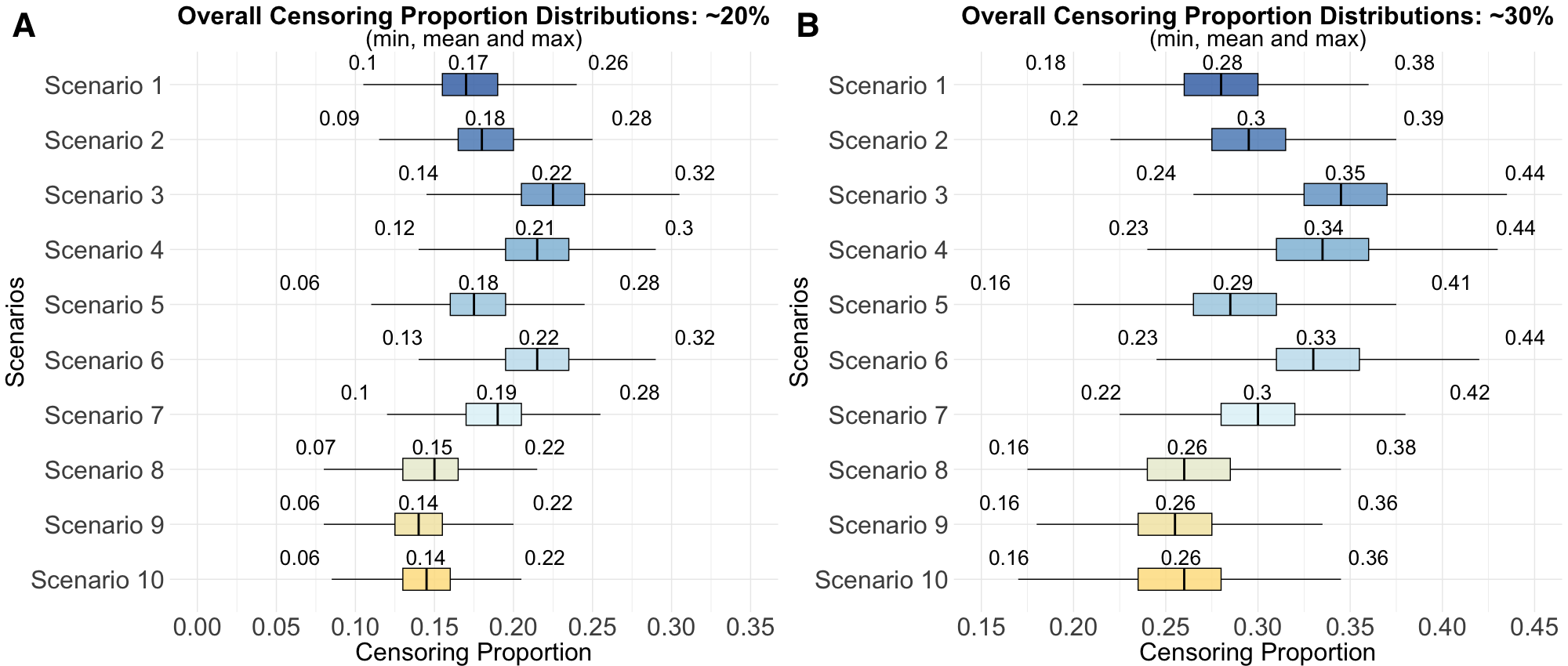}
    \caption{The boxplots of overall censoring proportion for scenarios 1 to 10 when the average overall censoring proportion among all scenarios is around (A) 20\% and (B) 30\%. The numbers aligned with each boxplot from left to right represent the minimum, mean and maximum of the overall censoring proportion among 500 simulation replicates.}
    \label{cen_overall_AB}
\end{figure}

\begin{figure}
    \centering
    \includegraphics[width = 1.0\textwidth]{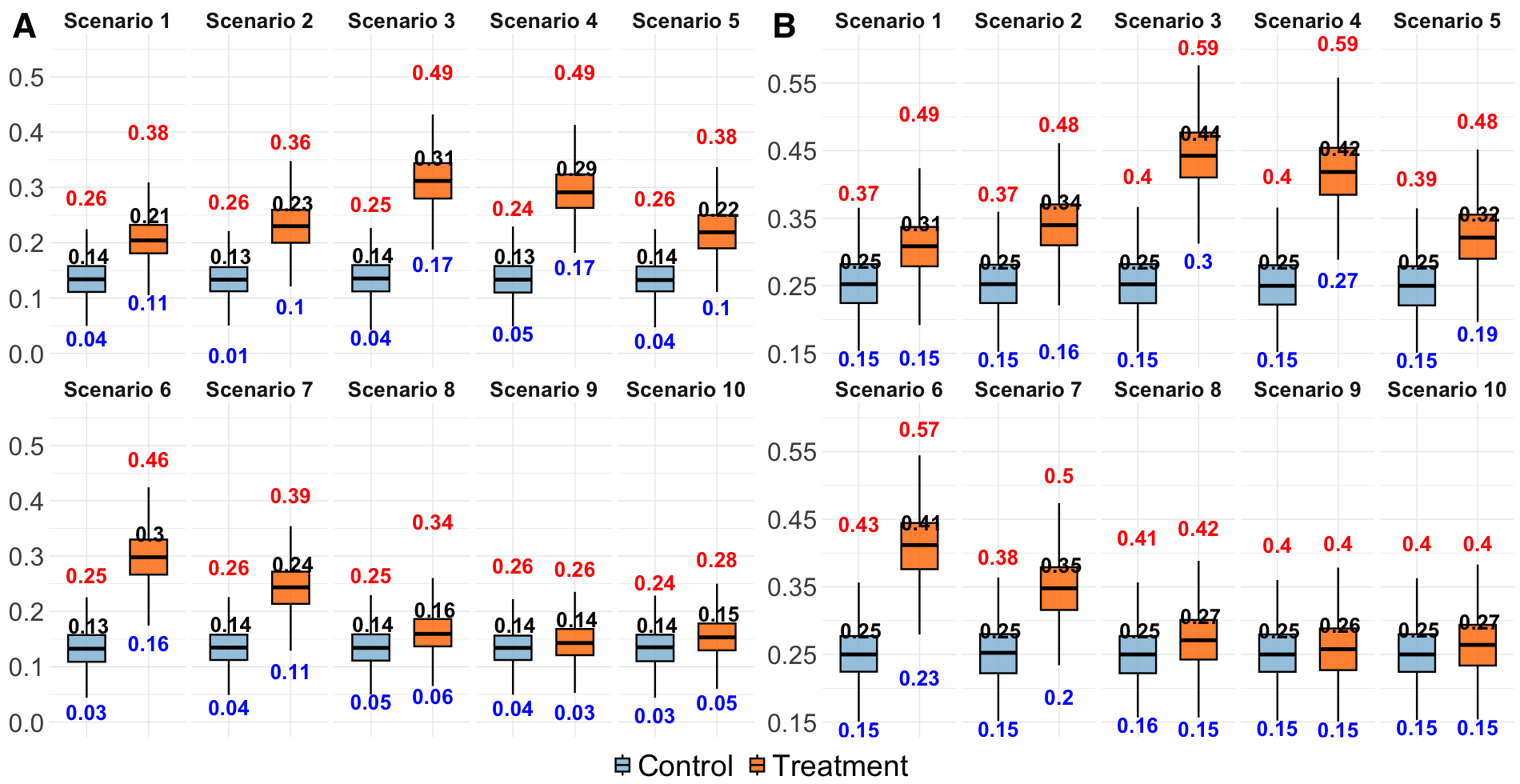}
    \caption{The boxplots of treatment (bright orange) and control (light blue) group censoring proportion for scenarios 1 to 10 when the average overall censoring proportion among all scenarios is around (A) 20\% and (B) 30\%. The numbers aligned with each boxplot from bottom to top represent the minimum (blue), mean (black) and maximum (red) of corresponding censoring proportion among 500 simulation replicates.}
    \label{cen_double_AB}
\end{figure}

We evaluated the performance of TADA under different strategies for censoring. For low to medium overall censoring proportions of 20\% and 30\%, Figure~\ref{cen_overall_AB} provides the distribution of the overall censoring rate for each scenario, while Figure~\ref{cen_double_AB} presents the censoring rates for the treatment and control groups across various scenarios. Tables~\ref{table-cen0.2} and~\ref{table-cen0.3} summarize key simulation results for both the censoring-adjusted and censoring-ignored cases across all scenarios. Specifically, each table compares the estimated bias and coverage relative to the pseudo-true hazard ratio. The illustrations of the overall censoring rate distribution, as well as the simulation results under relatively high overall censoring proportions of 40\% and 50\%, are provided in the Supplementary Materials S3. 

In general, TADA with censoring adjustments demonstrates improved efficiency in terms of bias and coverage compared to the unadjusted approach across various censoring scenarios. Compared to the true hazard ratio, the censoring-adjusted TADA consistently yields lower bias and improved coverage in most cases. As the difference in censoring rates between treatment and control groups grows (e.g., in Scenarios 3, 4, and 6), the benefits of the censoring adjustment become increasingly evident across all tested levels of overall censoring. 

At lower censoring levels, the advantages of the censoring-adjusted TADA approach are prominent. Specifically, with a 20\% censoring rate, the adjusted HR estimates exhibit lower bias than those unadjusted. For example, in Scenario 1, the bias associated with the censoring-adjusted method is 0.040 (95\% CI: –0.256, 0.449), which is less than half of the bias under the censoring-ignored method (0.114; 95\% CI: –0.205, 0.572) and with a narrower CI. This performance gap is further amplified in scenarios with greater censoring imbalances between treatment groups. In Scenario 3, the adjusted method yields a nearly unbiased point estimate, while the unadjusted estimate incurs a bias of eight times higher in magnitude. Similarly, in Scenario 4, the bias is reduced from 0.076 to 0.022, and in Scenario 6, from 0.086 to 0.023, when censoring is properly adjusted. The censoring-adjusted strategy also improves coverage. At 20\% overall censoring, nine out of ten scenarios attain coverage above 90\%, compared to six out of ten falling in the unadjusted cases. The improvement is more prominent when overall censoring increases to 30\%, though there is a slight decline in the absolute values of coverage.

When the overall censoring rate reaches 40\% and 50\%, both the censoring-adjusted and unadjusted estimators exhibit larger bias and wider confidence intervals due to severe data loss. Nevertheless, TADA with censoring adjustment continues to show substantially lower bias—often around 30\%–50\% less—compared to the unadjusted approach, and its coverage generally remains in a higher range. By contrast, coverage for the unadjusted estimator can drop well below 70\%, reaching as low as 50\% in certain cases. Full details, including scenario-specific hazard ratios, bias, and coverage, are provided in Supplementary Materials S3.

These findings confirm that although high overall censoring inevitably compromises estimation precision, accounting for censoring through proper adjustment is critical to preserving inferential validity. The censoring-adjusted TADA approach consistently outperforms the unadjusted method in both bias control and coverage maintenance, even under severe data loss conditions. Moreover, as shown in Figure~\ref{cen_double_AB} and S2, the censoring proportions between treatment and control never diverge to an extreme. This balanced design corresponds to many real-world scenarios where group-wise censoring differences remain moderate, ensuring TADA's adjustment results are applicable to commonly encountered settings.

\begin{landscape}
    
\begin{table}
\centering

\renewcommand{\arraystretch}{1.6}

\resizebox{1.35\textwidth}{!}{%

\begin{tabular}{ccccccccc}
\toprule
\textbf{Scenario} & \textbf{HR (95\% CI): Censor} & \textbf{HR (95\% CI): Non-censor} & \textbf{HR (95\% CI): True} & \textbf{Bias (95\% CI): Censor} & \textbf{Bias (95\% CI): Non-censor} & \textbf{Coverage: Censor} & \textbf{Coverage: Non-Censor} \\
\midrule
1  & 0.858 (0.575, 1.295) & 0.932 (0.624, 1.411) & 0.818 (0.811, 0.826) & 0.040 (–0.256, 0.449) & 0.114 (–0.205, 0.572) & 0.936 & 0.910 \\
2  & 0.797 (0.536, 1.191) & 0.865 (0.580, 1.299) & 0.753 (0.746, 0.760) & 0.044 (–0.222, 0.365) & 0.112 (–0.173, 0.486) & 0.938 & 0.904 \\
3  & 0.598 (0.400, 0.905) & 0.647 (0.429, 0.996) & 0.591 (0.586, 0.597) & 0.007 (–0.208, 0.260) & 0.056 (–0.183, 0.351) & 0.934 & 0.932 \\
4  & 0.645 (0.430, 0.964) & 0.699 (0.464, 1.058) & 0.623 (0.617, 0.629) & 0.022 (–0.192, 0.328) & 0.076 (–0.171, 0.428) & 0.928 & 0.904 \\
5  & 0.828 (0.555, 1.244) & 0.902 (0.604, 1.363) & 0.789 (0.782, 0.796) & 0.040 (–0.267, 0.430) & 0.113 (–0.214, 0.551) & 0.936 & 0.878 \\
6  & 0.647 (0.433, 0.974) & 0.710 (0.473, 1.083) & 0.624 (0.618, 0.630) & 0.023 (–0.190, 0.319) & 0.086 (–0.145, 0.432) & 0.948 & 0.898 \\
7  & 0.771 (0.516, 1.164) & 0.844 (0.563, 1.281) & 0.730 (0.724, 0.737) & 0.041 (–0.229, 0.382) & 0.113 (–0.186, 0.465) & 0.944 & 0.896 \\
8  & 1.014 (0.694, 1.535) & 1.059 (0.724, 1.596) & 0.952 (0.943, 0.960) & 0.063 (–0.266, 0.503) & 0.107 (–0.252, 0.590) & 0.930 & 0.914 \\
9  & 1.058 (0.722, 1.580) & 1.079 (0.741, 1.600) & 0.990 (0.981, 0.999) & 0.068 (–0.294, 0.540) & 0.089 (–0.285, 0.553) & 0.908 & 0.910 \\
10 & 0.923 (0.622, 1.384) & 0.959 (0.654, 1.425) & 0.796 (0.789, 0.803) & 0.127 (–0.179, 0.502) & 0.163 (–0.154, 0.551) & 0.892 & 0.846 \\
\bottomrule
\end{tabular}%
}

\caption{The simulation results of scenarios 1 to 10 when the overall censoring proportion is around 20\%. The columns from left to right represent scenario index; estimated hazard ratio and 95\% CI when TADA adjusts censoring; estimated hazard ratio and 95\% CI when TADA ignores censoring; pseudo-true hazard ratio and 95\% CI; bias and 95\% CI between censoring-adjusted strategy and pseudo-true hazard ratio; bias and 95\% CI between censoring-ignored strategy and pseudo-true hazard ratio; coverage of pseudo-true hazard ratio when adopting censoring-adjusted strategy; coverage of pseudo-true hazard ratio when adopting censoring-ignored strategy.}

\label{table-cen0.2}

\centering
\resizebox{1.35\textwidth}{!}{%

\begin{tabular}{ccccccccc}
\toprule
\textbf{Scenario} & \textbf{HR (95\% CI): Censor} & \textbf{HR (95\% CI): Non-censor} & \textbf{HR (95\% CI): True} & \textbf{Bias (95\% CI): Censor} & \textbf{Bias (95\% CI): Non-censor} & \textbf{Coverage: Censor} & \textbf{Coverage: Non-Censor} \\
\midrule
1  & 0.935 (0.613, 1.463) & 1.062 (0.697, 1.659) & 0.818 (0.811, 0.826) & 0.116 (–0.225, 0.647) & 0.244 (–0.143, 0.817) & 0.898 & 0.780 \\
2  & 0.869 (0.571, 1.350) & 0.980 (0.644, 1.527) & 0.753 (0.746, 0.760) & 0.116 (–0.185, 0.535) & 0.227 (–0.114, 0.693) & 0.902 & 0.764 \\
3  & 0.642 (0.418, 1.009) & 0.722 (0.466, 1.155) & 0.591 (0.586, 0.597) & 0.051 (–0.197, 0.337) & 0.131 (–0.157, 0.467) & 0.934 & 0.868 \\
4  & 0.697 (0.455, 1.083) & 0.788 (0.511, 1.238) & 0.623 (0.617, 0.629) & 0.074 (–0.178, 0.432) & 0.165 (–0.117, 0.598) & 0.932 & 0.822 \\
5  & 0.896 (0.588, 1.401) & 1.020 (0.668, 1.595) & 0.789 (0.782, 0.796) & 0.107 (–0.249, 0.577) & 0.232 (–0.160, 0.762) & 0.900 & 0.792 \\
6  & 0.699 (0.456, 1.096) & 0.805 (0.523, 1.273) & 0.624 (0.618, 0.630) & 0.075 (–0.167, 0.410) & 0.181 (–0.088, 0.585) & 0.916 & 0.792 \\
7  & 0.837 (0.548, 1.308) & 0.958 (0.626, 1.505) & 0.730 (0.724, 0.737) & 0.106 (–0.202, 0.523) & 0.228 (–0.145, 0.691) & 0.902 & 0.778 \\
8  & 1.055 (0.701, 1.646) & 1.133 (0.754, 1.755) & 0.952 (0.943, 0.960) & 0.103 (–0.284, 0.558) & 0.181 (–0.231, 0.730) & 0.924 & 0.876 \\
9  & 1.069 (0.709, 1.650) & 1.112 (0.742, 1.704) & 0.990 (0.981, 0.999) & 0.079 (–0.324, 0.551) & 0.121 (–0.304, 0.607) & 0.924 & 0.902 \\
10 & 0.957 (0.629, 1.481) & 1.013 (0.672, 1.555) & 0.796 (0.789, 0.803) & 0.161 (–0.192, 0.604) & 0.217 (–0.156, 0.713) & 0.864 & 0.788 \\
\bottomrule
\end{tabular}%

}
\caption{The simulation results of scenarios 1 to 10 when the overall censoring proportion is around 30\%. The columns from left to right represent scenario index; estimated hazard ratio and 95\% CI when TADA adjusts censoring; estimated hazard ratio and 95\% CI when TADA ignores censoring; pseudo-true hazard ratio and 95\% CI; bias and 95\% CI between censoring-adjusted strategy and pseudo-true hazard ratio; bias and 95\% CI between censoring-ignored strategy and pseudo-true hazard ratio; coverage of pseudo-true hazard ratio when adopting censoring-adjusted strategy; coverage of pseudo-true hazard ratio when adopting censoring-ignored strategy.}

\label{table-cen0.3}
\end{table}

\end{landscape}

\section{Sensitivity Analysis}
\label{sec6}
\setstretch{1.3}

To fully evaluate the proposed TADA method, we perform comprehensive sensitivity analyses in several aspects, including the truncation strategies of the final weights, the sample size of the study data, and possible abnormal values in the target population, corresponding to the various elements in the process that may affect the performance of TADA. Without loss of generality, the following simulations are conducted with the relevant settings of Scenario 1 in Section 4.1 as a baseline, with overall censoring rate of 20\%. Each case simulates 500 replicates and each replicate runs 200 bootstrap resamples.

\subsection{Final Weights Truncation Strategy}

To assess the impact of final weight truncation on the performance of TADA, we vary the quantile cutoffs applied to the final weights, ranging from 80\% to 99\%, alongside a no-truncation baseline. As shown in Table~\ref{tab:truncation_sensitivity}, in the censoring-ignored setting where only participation weights are applied, the bias decreases as the truncation becomes stricter, and the confidence intervals correspondingly narrow. This reflects the dominant influence of extreme weights derived from covariate imbalance, which are progressively mitigated under tighter cutoffs. In the censoring-adjusted setting where the final weights incorporate both participation and censoring components, the trend of bias exhibits a convex pattern: it is reduced at moderate cutoffs but may increase under excessively loose or tight truncation. This may arise from competing effects—retaining large IPCW values inflates variance and bias, while aggressive truncation may underweight censored individuals and induce instability. Despite this, the censoring-adjusted approach maintains lower bias and higher coverage across all settings, indicating a consistent robustness. These results suggest that a proper truncation is a useful safeguard to improve estimator reliability in practice.

\subsection{Source Sample Size and Bias Source}

To better understand the source and behavior of estimation bias of TADA, we increase the source sample size stepwise while holding all other aspects of the simulation design fixed. Specifically, we vary the source sample size \(N_{\text{s}} \in \{300, 500, 800, 1000\}\), while maintaining the same baseline covariate distributions, censoring mechanism, and final weight truncation threshold across all settings. According to Table~\ref{tab:sample_size}, the bias decreases gradually from \(N_\textit{s}=300\) to \(1000\) and the confidence intervals all include zero, suggesting modest bias. Larger datasets may reduce this bias further, but we limit \(N_\textit{s}\) to \(1000\) considering the computational cost under time-varying weighting. The coverage results of censoring-adjusted cases are robust with the increase of source sample size, which supports the consistency of TADA. The close alignment between Monte Carlo standard deviations and bootstrap standard errors indicates relatively accurate variance estimation; hence, some coverage shortfalls shown in Section~\ref{sec5} may reflect small-sample bias or censoring effects rather than variance underestimation. Alternative resampling schemes, such as stratified or weighted bootstrap, may be more suitable than the naive bootstrap in large-sample settings with substantial censoring.

\subsection{Abnormal Values in the Target Population}

To evaluate the robustness of the TADA in the presence of potential abnormal values in the target population, we conduct a series of scenarios designed to mimic practical data irregularities that may compromise the reliability of sample moments. As shown in Table~\ref{tab:abv_definition}, we consider seven predefined scenarios, each introducing distinct forms of abnormality into the target population data. In the clean baseline scenario (Type 1), no modification is applied. In Types 2 and 3, we introduce binary noise by randomly selecting 1\% or 5\% of individuals and flipping the values of the binary covariate $X_2$, simulating potential misclassification or data entry errors. Types 4 and 5 simulate extreme outliers in the continuous covariate $X_3$ by adding several times of original standard deviations to 1\% or 5\% of randomly selected observations, capturing heavy-tailed or contaminated measurement scenarios. Types 6 and 7 combine both mentioned mechanisms, simultaneously introducing flips in $X_2$ and extreme values in $X_3$ at corresponding proportions.

\begin{table}[ht]
  \centering
  \renewcommand{\arraystretch}{1.4}
  \resizebox{\textwidth}{!}{%
    \begin{tabular}{cccc}
      \toprule
      \textbf{ABV Type}
        & \textbf{Proportion of Flipped in $X_2$}
        & \textbf{Proportion of Extreme Values in $X_3$}
        & \textbf{Extreme Multiplier of $X_3$} \\
      \midrule
      1 & --   & --   & -- \\
      2 & 1\%  & --   & -- \\
      3 & 5\%  & --   & -- \\
      4 & --   & 1\%  & 3x \\
      5 & --   & 5\%  & 3x \\
      6 & 1\%  & 1\%  & 3x \\
      7 & 5\%  & 5\%  & 3x \\
      \bottomrule
    \end{tabular}%
  }
  \caption{Abnormal value pattern types in the simulated target population. ``Proportion of Flipped'' means the proportion of binary variable values that appear to have flip-flopped from their true values. ``Proportion of Extreme Values'' means the proportion of continuous variable values that are extremely skewed compared to the overall distribution. ``Extreme Multiplier'' means the multiplier used to construct the continuous variable outliers. The hyphen indicates the corresponding parameter setting is not in effect. ABV Abnormal value.}
  \label{tab:abv_definition}
\end{table}

According to Table~\ref{tab:abv_collapsed}, when existing outliers in the target population, the TADA estimates show a certain degree of increase in bias and widening of the confidence intervals as the form of the outliers became more complex from a single type to the presence of a superposition of both flipped and polarized values. However, the degradation of performance is limited and the coverage and bias levels remain within acceptable limits, demonstrating overall robustness. Overall, TADA is able to control bias when faced with moderate extreme values in the target population data.

\begin{landscape}

\begin{table}

\renewcommand{\arraystretch}{1.6}

\centering
\resizebox{1.35\textwidth}{!}{%
\begin{tabular}{cccccccc}
\toprule
\textbf{Trunc.\ cutoff} 
    & \textbf{HR (95\% CI): Censor} 
    & \textbf{HR (95\% CI): Non-censor} 
    & \textbf{HR (95\% CI): True} 
    & \textbf{Bias (95\% CI): Censor} 
    & \textbf{Bias (95\% CI): Non-censor} 
    & \textbf{Coverage: Censor} 
    & \textbf{Coverage: Non-Censor} \\
\midrule
80\%          
& 0.886 (0.628, 1.259) & 0.925 (0.657, 1.317) & 0.818 (0.811, 0.826) 
& 0.068 (–0.191, 0.439) & 0.107 (–0.170, 0.492) 
& 92.00\% & 88.80\% \\

90\%          
& 0.869 (0.600, 1.266) & 0.926 (0.640, 1.355) & 0.818 (0.811, 0.826) 
& 0.050 (–0.221, 0.448) & 0.108 (–0.180, 0.509) 
& 93.20\% & 90.80\% \\

95\%          
& 0.858 (0.575, 1.295) & 0.932 (0.624, 1.411) & 0.818 (0.811, 0.826) 
& 0.040 (–0.256, 0.449) & 0.114 (–0.205, 0.572) 
& 93.60\% & 91.00\% \\

99\%          
& 0.849 (0.547, 1.353) & 0.950 (0.609, 1.508) & 0.818 (0.811, 0.826) 
& 0.031 (–0.294, 0.524) & 0.131 (–0.219, 0.644) 
& 94.00\% & 90.20\% \\

No Truncation 
& 0.861 (0.545, 1.388) & 0.973 (0.617, 1.546) & 0.818 (0.811, 0.826) 
& 0.043 (–0.307, 0.631) & 0.155 (–0.240, 0.764) 
& 92.60\% & 86.80\% \\
\bottomrule
\end{tabular}%
}
\caption{The simulation results of various truncation strategies for $w^\textbf{final}$. The columns from left to right represent truncation cutoff values of $w^\textbf{final}$; estimated hazard ratio and 95\% CI when TADA adjusts censoring; estimated hazard ratio and 95\% CI when TADA ignores censoring; pseudo-true hazard ratio and 95\% CI; bias and 95\% CI between censoring-adjusted strategy and pseudo-true hazard ratio; bias and 95\% CI between censoring-ignored strategy and pseudo-true hazard ratio; coverage of pseudo-true hazard ratio when adopting censoring-adjusted strategy; coverage of pseudo-true hazard ratio when adopting censoring-ignored strategy. All the other settings in the simulations are identical.}
\label{tab:truncation_sensitivity}

\centering
\resizebox{1.4\textwidth}{!}{%
\begin{tabular}{cccccccccc}
\toprule
\textbf{$N_{\textit{s}}$}
  & \textbf{Bias (95\% CI): Censor}
  & \textbf{Bias (95\% CI): Non-censor}
  & \textbf{Coverage: Censor} 
  & \textbf{Coverage: Non-Censor}
  & \textbf{Monte Carlo SD: Censor}
  & \textbf{Bootstrap SE: Censor} 
  & \textbf{Monte Carlo SD: Non-censor}
  & \textbf{Bootstrap SE: Non-censor} \\
\midrule
300  
& 0.027 (–0.197, 0.339) 
& 0.102 (–0.143, 0.435) 
& 94.00\% 
& 89.60\% 
& 0.13\% 
& 0.14\% 
& 0.14\% 
& 0.15\% \\

500 
& 0.026 (–0.162, 0.259) 
& 0.099 (–0.108, 0.353)
& 91.80\% 
& 81.80\% 
& 0.11\% 
& 0.10\% 
& 0.12\% 
& 0.11\% \\

800  
& 0.024 (–0.121, 0.176) 
& 0.099 (–0.066, 0.266)
& 95.40\% 
& 77.40\% 
& 0.08\% 
& 0.08\%
& 0.08\% 
& 0.09\% \\

1000 
& 0.021 (–0.104, 0.153) 
& 0.096 (–0.050, 0.244) 
& 93.60\% 
& 73.40\% 
& 0.07\% 
& 0.07\% 
& 0.08\% 
& 0.08\% \\

\bottomrule
\end{tabular}%
}
\caption{The simulation results of various source sample size $N_\textit{s}$ with corresponding Monte Carlo SD and Bootstrap SE. The columns from left to right represent source sample size $N_\textit{s}$; bias and 95\% CI between censoring-adjusted strategy and pseudo-true hazard ratio; bias and 95\% CI between censoring-ignored strategy and pseudo-true hazard ratio; coverage of pseudo-true hazard ratio when adopting censoring-adjusted strategy; coverage of pseudo-true hazard ratio when adopting censoring-ignored strategy; Monte Carlo SD for the censoring-adjusted strategy; Bootstrap SE for the censoring-adjusted strategy; Monte Carlo SD for the censoring-ignored strategy; Bootstrap SE for the censoring-ignored strategy. All the other settings in the simulations are identical. SD Standard deviation. SE Standard error.}
\label{tab:sample_size}

\resizebox{1.35\textwidth}{!}{%
\begin{tabular}{cccccccc}
\toprule
\textbf{ABV Type}
  & \textbf{HR (95\% CI): Censor}
  & \textbf{HR (95\% CI): Non-censor}
  & \textbf{HR (95\% CI): True}
  & \textbf{Bias (95\% CI): Censor}
  & \textbf{Bias (95\% CI): Non-censor}
  & \textbf{Coverage: Censor}
  & \textbf{Coverage: Non-Censor} \\
\midrule
1 
  & 0.858 (0.575, 1.296) & 0.932 (0.625, 1.412) & 0.818 (0.811, 0.826)
  & 0.039 (–0.253, 0.437) & 0.114 (–0.210, 0.543)
  & 92.80\% & 91.40\% \\
2 
  & 0.858 (0.574, 1.295) & 0.932 (0.624, 1.410) & 0.818 (0.811, 0.826)
  & 0.039 (–0.251, 0.439) & 0.114 (–0.207, 0.559)
  & 93.20\% & 91.80\% \\
3 
  & 0.857 (0.573, 1.294) & 0.931 (0.623, 1.408) & 0.818 (0.811, 0.826)
  & 0.038 (–0.258, 0.428) & 0.113 (–0.206, 0.543)
  & 94.20\% & 92.00\% \\
4 
  & 0.862 (0.561, 1.338) & 0.939 (0.611, 1.465) & 0.818 (0.811, 0.826)
  & 0.044 (–0.257, 0.452) & 0.121 (–0.206, 0.595)
  & 93.80\% & 92.40\% \\
5 
  & 0.881 (0.493, 1.673) & 0.971 (0.539, 1.737) & 0.818 (0.811, 0.826)
  & 0.063 (–0.320, 0.629) & 0.152 (–0.279, 0.791)
  & 95.00\% & 93.40\% \\
6 
  & 0.860 (0.561, 1.335) & 0.937 (0.611, 1.463) & 0.818 (0.811, 0.826)
  & 0.042 (–0.272, 0.448) & 0.119 (–0.229, 0.570)
  & 93.60\% & 93.20\% \\
7 
  & 0.882 (0.493, 1.680) & 0.972 (0.535, 1.748) & 0.818 (0.811, 0.826)
  & 0.063 (–0.323, 0.664) & 0.153 (–0.269, 0.846)
  & 94.80\% & 93.20\% \\
\bottomrule
\end{tabular}%
}
\caption{The simulation results of various abnormal value patterns in the target population. The columns from left to right represent abnormal value types in Table~\ref{tab:abv_definition}; estimated hazard ratio and 95\% CI when TADA adjusts censoring; estimated hazard ratio and 95\% CI when TADA ignores censoring; pseudo-true hazard ratio and 95\% CI; bias and 95\% CI between censoring-adjusted strategy and pseudo-true hazard ratio; bias and 95\% CI between censoring-ignored strategy and pseudo-true hazard ratio; coverage of pseudo-true hazard ratio when adopting censoring-adjusted strategy; coverage of pseudo-true hazard ratio when adopting censoring-ignored strategy. All the other settings in the simulations are identical. ABV Abnormal value.}
\label{tab:abv_collapsed}

\end{table}

\end{landscape}

\section{Real Case Study}
\label{sec7}
\setstretch{1.3}

To demonstrate the real-world applicability and practical value of the proposed TADA method, we conduct an application that transports survival outcomes from an RCT with censoring to a real-world population with only aggregate-level data. The source dataset is derived from the SQUIRE trial (NCT00981058)~\citep{thatcher2015necitumumab}, a phase III, multi-center, open-label RCT evaluating the addition of necitumumab to gemcitabine and cisplatin chemotherapy in patients with stage IV squamous non-small-cell lung cancer (NSCLC). Patients are randomized in a 1:1 ratio to receive either necitumumab plus chemotherapy or chemotherapy alone. We use the publicly available IPD from the Project Data Sphere platform~\citep{green2015project}, which contained only patients randomized to the control (gemcitabine/cisplatin) arm ($N_\textit{s}$ = 548). Available source sample data includes overall survival time, censoring status, and relevant baseline covariates such as age, sex, ethnic origin, smoking history and Eastern Cooperative Oncology Group (ECOG) performance status. The overall censoring proportion is around 19\%.

The target population is defined based on a population-based cohort study in Ontario, Canada, reported by Seung et al.~\citep{seung2020real}. This registry-based dataset includes patients diagnosed with stage IV NSCLC who received first-line chemotherapy between 2007 and 2015. We focus on squamous histology patients to align with the SQUIRE inclusion criteria ($N_\textit{t} = \text{2,056}$). Only summary-level statistics are reported for this subgroup: the average age is 72 years, and 65.1\% of patients are male. In contrast, the SQUIRE trial’s control arm is younger (mean age: 61.7 years) and has a higher proportion of males (83.6\%). The baseline characteristics differences between source and target populations are summarized in Table~\ref{tab:real_covs}.

\begin{table}[htbp]
\renewcommand{\arraystretch}{1.4}
\centering
\caption{Baseline characteristics of the source population in SQUIRE trial and the target population in Ontario study.}
\label{tab:real_covs}
\begin{tabular}{lcc}
\toprule
 & \textbf{Source (SQUIRE)} & \textbf{Target (Ontario)} \\
\midrule
Sample size             & 548    & 2{,}056\\
Age (mean in years)     & 61.7   & 72.0 \\
Male (\%)               & 83.6   & 65.1 \\
\bottomrule
\end{tabular}
\end{table}

The estimand of interest is the marginal survival function over time, with particular focus on the estimated median OS and corresponding 95\% confidence intervals. The survival function is estimated using weighted Kaplan–Meier methods under various adjustments on covariate imbalancing and censoring. We apply age and sex as the population matching covariates while adding ethnic origin, smoking history and ECOG performance status when estimating the censoring weights. In particular, we consider four adjustment strategies as follows:
\begin{itemize}
    \item \textbf{Original:} The standard Kaplan–Meier estimator applied directly to the source IPD, without any weighting, serving as a naive benchmark that ignores both covariate imbalances and censoring-related bias.
    \item \textbf{Original with Censoring Adjusted:} An estimator that applies censoring weights to the source data in order to mitigate bias due to incomplete follow-up, but without accounting for population-level covariate mismatches.
    \item \textbf{Transported without Censoring Adjusted:} A weighted Kaplan–Meier estimator incorporating participation weights to align covariate distributions between the source and target populations, while disregarding potential bias introduced by censoring.
    \item \textbf{TADA (Transported with Censoring Adjusted):} The proposed TADA method combining both participation weights and censoring weights, simultaneously correcting for censoring and aligning the source population with the target population on the key covariates.
\end{itemize}

\begin{figure}[htbp]
\centering
\includegraphics[width=1.0\textwidth]{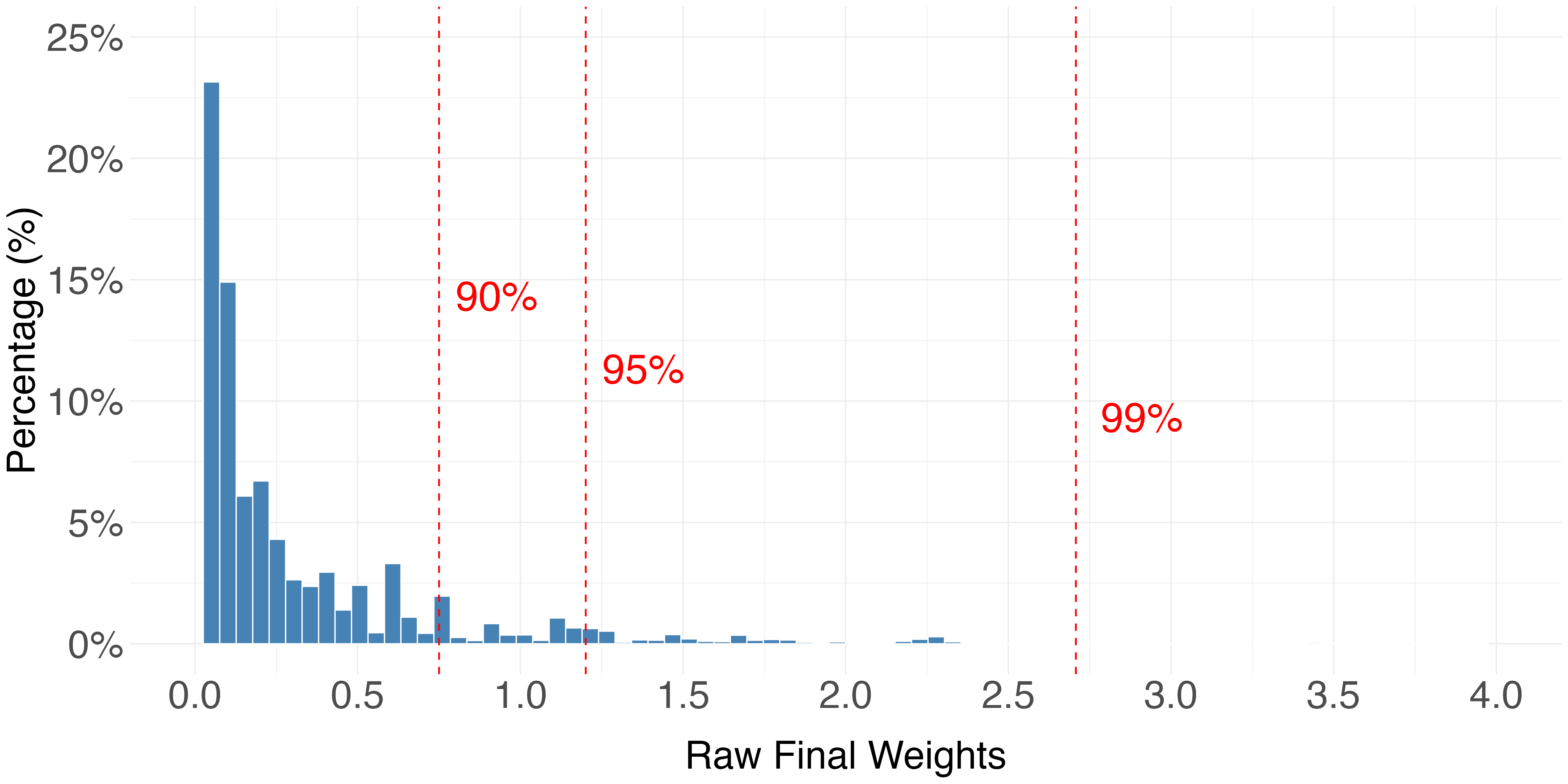}
\caption{An illustration of the final weights truncation threshold decision. The horizontal axis represents the value of raw final weights. The vertical axis indicates the percentage of observations in each weight bin, relative to the total sample. Blue bars depict the distribution of raw final weights. Red vertical dashed lines from left to right indicate the candidate thresholds at the $90^\text{th}$, $95^\text{th}$ and $99^\text{th}$ percentiles, respectively.}
\label{fig:trunc_judge}
\end{figure}

To obtain the truncated final weights, we truncate the raw final weights at a preferred threshold quantile based on their distribution after assessments. Figure~\ref{fig:trunc_judge} illustrates this process: for the shown raw final weights, applying truncation at the 95\% quantile could be the optimal choice for current scenario to keep most of information and avoid the extreme weight candidates. We apply the 95\% quantile as the truncation threshold for the introduced adjustments in the real case study, supported by assessment results.

\begin{figure}[htbp]
\centering
\includegraphics[width=1.0\textwidth]{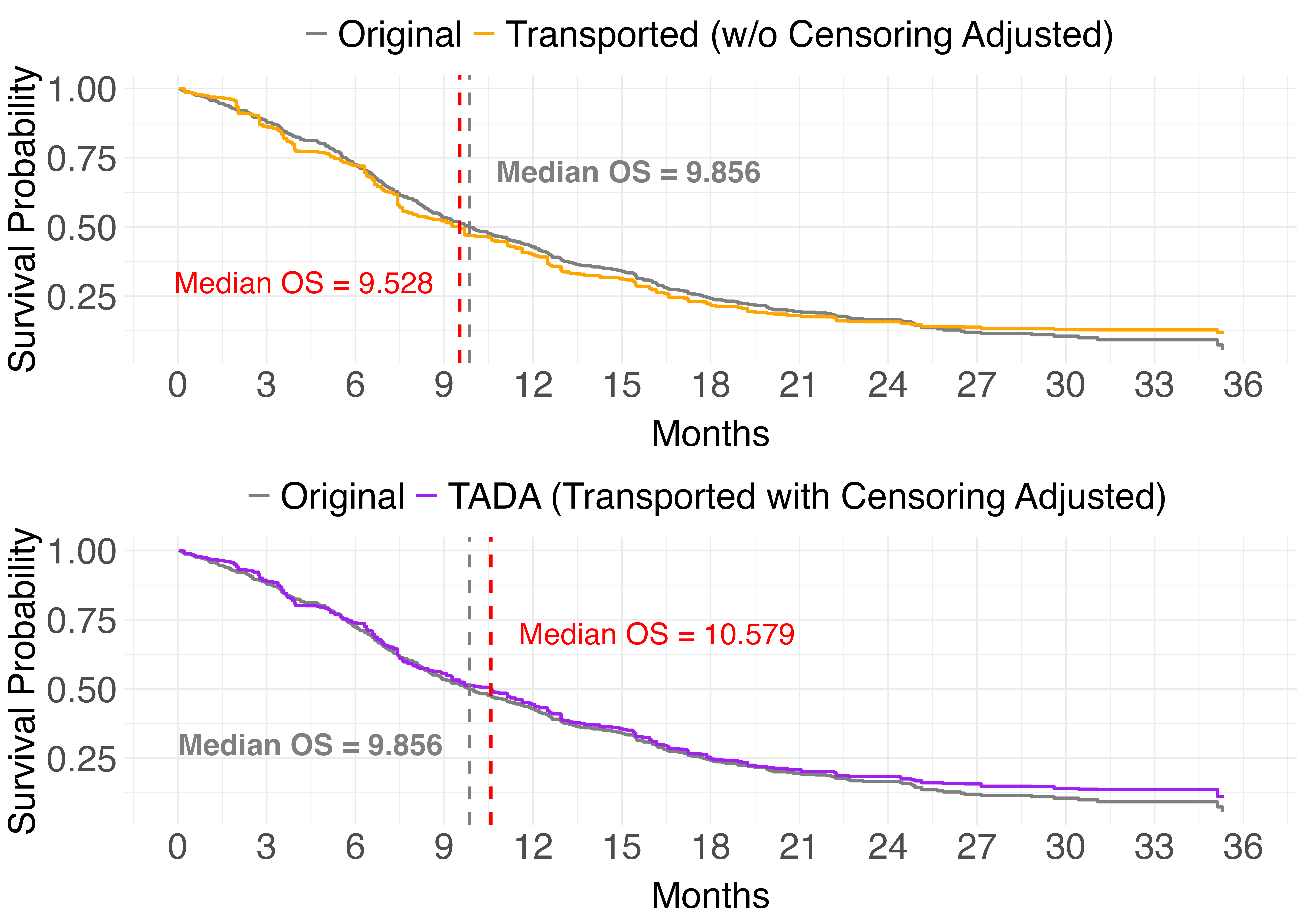}
\caption{Original Kaplan--Meier curve and Transported Kaplan--Meier curves under various adjustments. 
\textbf{Top panel:} Original naive K-M curve vs. Transported K-M curve that adjusted for covariate imbalance and without censoring adjustment.  
\textbf{Bottom panel:} Original naive K-M curve vs. Transported K-M curve fully adjusted by TADA, incorporating both censoring and covariate imbalance adjustments.  
Dashed vertical lines in grey indicate the estimated median OS for the naive curve. Dashed vertical lines in red indicate the estimated median OS for the transported curves.}
\label{fig:KM-results-pairs}
\end{figure}

Figure~\ref{fig:KM-results-pairs} displays the comparison between the unadjusted naive Kaplan–Meier curve and two various weighted Kaplan–Meier curves, representing the transported cases with and without censoring adjustment respectively. Table~\ref{tab:real_OS} reports the corresponding estimated median OS and 95\% confidence intervals. Table~\ref{tab:surv_prob} summarizes the estimated survival probabilities at every 6 months under different adjustments. The original median OS without any adjustment is 9.856 months, while the transported median OS without censoring adjustment is 9.528 months. Adjusting only for censoring leads to a close median OS as the naive case of 9.889 months. The fully adjusted TADA estimator, simultaneously correcting for both covariate imbalance and censoring bias, yields an estimated median OS of 10.579 months and a narrower confidence interval compared to that of the transported without censoring adjustment result. These estimates illustrate the differences arising from adjustments for censoring and covariate distribution imbalance when transporting clinical trial outcomes to real-world populations, which may lead to different clinical conclusions. Seung et al.~\citep{seung2020real} also emphasized that real-world cohorts often include older patients compared to those enrolled in clinical trials. As such, the ability of TADA to account for population differences and censoring enhances the external validity of the transported estimates. Due to limited data availability, this case study incorporated only the control arm datasets and limited set of covariates. The inclusion of additional information such as the treatment arm data and more key covariates would yield more clinically relevant survival estimates.

\begin{table}[htbp]
\renewcommand{\arraystretch}{1.4}
\centering
\caption{Estimated median overall survival (OS) time and 95\% confidence intervals under different adjustments.}
\resizebox{\textwidth}{!}{
\begin{tabular}{lccc}
\toprule
\textbf{} & \textbf{Median OS (months)} & \textbf{95\% CI (Lower)} & \textbf{95\% CI (Upper)} \\
\midrule
Original                              & 9.856 &  8.903 & 11.138 \\
Original (with Censoring Adjusted)    & 9.889 &  8.903 & 11.138 \\
Transported (w/o Censoring Adjusted)  & 9.528 &  7.425 & 11.696 \\
TADA fully adjusted                   & 10.579 & 8.608 & 12.485 \\
\bottomrule
\end{tabular}
}
\label{tab:real_OS}
\end{table}

\begin{table}[htbp]
\renewcommand{\arraystretch}{1.4}
\centering
\caption{Estimated survival probabilities at every 6 months under different adjustments.}
\label{tab:survival_probs}
\resizebox{\textwidth}{!}{
\begin{tabular}{ccccc}
\toprule
\textbf{Month} 
    & \textbf{Original} 
    & \textbf{Original (with Censoring Adjusted)} 
    & \textbf{Transported (w/o Censoring Adjusted)} 
    & \textbf{TADA} \\
\midrule
6   & 0.723 & 0.723 & 0.721 & 0.738 \\
12  & 0.428 & 0.428 & 0.401 & 0.443 \\
18  & 0.243 & 0.242 & 0.222 & 0.254 \\
24  & 0.165 & 0.165 & 0.157 & 0.183 \\
30  & 0.106 & 0.106 & 0.129 & 0.141 \\
36  & 0.055 & 0.056 & 0.114 & 0.107 \\
\bottomrule
\end{tabular}
}
\label{tab:surv_prob}
\end{table}

Table~\ref{tab:real_OS} and~\ref{tab:surv_prob} show that, compared with the naive case, the trend of weighted survival curve remains similar when only adjusted with censoring weights. This consistency results from the fact that few observations are censored in the first 9 months, leading to the estimated censoring weights concentrate near 1 and the Kaplan–Meier estimate largely unaffected. Only in the far right tail of the distribution (i.g., beyond 18 months) does censoring become more frequent, at which point censoring weights exert a modest influence on the estimated survival curve. Figure~\ref{fig:diagnosis} visually demonstrates this phenomenon.

\begin{figure}[htbp]
\centering
\includegraphics[width=1.0\textwidth]{diagnosis.png}
\caption{Distribution of overall survival times by event status when only conduct censoring adjustment for the source SQUIRE population. The grey bars represent the status of event. The navy blue bars represents the status of censor. The red dashed line indicates the value of median OS.}
\label{fig:diagnosis}
\end{figure}

\section{Discussion}
\label{sec8}
\setstretch{1.3}

In this paper, we demonstrate that the proposed TADA method provides improvements in bias control and coverage across a range of censoring scenarios, showing effectiveness under low to moderate censoring proportions. These findings support the value of TADA in addressing censoring bias, with its advantage over unadjusted methods persisting even as overall censoring increases, albeit with some reduction in performance. The case study between SQUIRE and Ontario population provides an illustration of TADA's value in real application. Overall, these results suggest that TADA structurally contributes handling with information loss due to censoring, making it a valuable transportability analysis method in scenarios where IPDs are unavailable for the target population and where source data are disturbed by right-censoring.

A core advantage of the proposed TADA method lies in its flexible weighting-by-a-product structure, with each term specifically designed to adjust for different sources of confounding. This makes TADA highly adaptable to complex application scenarios. In current cases in TADA, the participation weights, derived via the MoM, can address the challenges of working with AgD for the target population for transportability analysis. In this study, we explored TADA's ability to transport survival outcomes under covariate-relevant censoring, which addresses the current gap in transportability methods which have generally been limited to binary and continuous outcomes and require IPD. 

In scenarios with modest censoring rates, the TADA method with an adjustment for censoring exhibits robust performance with low bias and high coverage. This effectiveness can be attributed to the lower level of information loss in low-censoring contexts, allowing TADA’s inverse probability weighting to make more precise adjustments. The stability of TADA’s performance in these scenarios highlights its ability to achieve accurate effect estimates under conditions of limited censoring, where the method leverages available data efficiently. In scenarios with a higher overall censoring proportion, TADA demonstrates performance close to its practical limit, with a decline in bias control and coverage. Increased censoring results in a higher loss of information which in turn reduces the sample size available for reliable adjustments. Higher censoring may result in more extreme weights and introduce greater variability in bias control. This accumulation of censoring-induced bias and weight instability creates challenges for maintaining accuracy and coverage under high-censoring conditions. Nonetheless, censoring-adjusted TADA outperforms the unadjusted approach in general.

In scenarios where there was a larger disparity in censoring proportions between treatment and control groups, TADA shows lower bias estimates. This phenomenon suggests that TADA may potentially benefit from this unique data characteristic. When one group has lower censoring, the more complete data could allow TADA’s weighting mechanism to draw on relatively stable information, which may enhance the reliability of adjustments. This natural heterogeneity might provide a form of implicit stratification, helping TADA balance distinct group effects with greater robustness. Such conditions could, in theory, offer additional stability in TADA’s performance by leveraging the more consistent information from less-censored data. However, as overall censoring increases, this differential effect diminishes, indicating that higher censoring rates may challenge TADA’s capacity to control bias across varying group censoring proportions. In our simulation study, we adopted the nonparametric bootstrap to estimate standard errors and coverage. This choice was driven by the method’s flexibility and its compatibility with the complex, non-closed-form weighting scheme in TADA. While bootstrapping is widely used in survival analysis and performs well under low to moderate censoring, it may become less reliable when substantial information is lost due to heavy censoring, which could partially explain the reduced coverage observed in some scenarios with 30\% or higher censoring. We note, however, that the primary purpose of our simulation study was to evaluate performance trends in bias control and coverage across scenarios, rather than to identify optimal variance estimation procedures under all conditions. Alternative inference strategies, such as robust sandwich variance estimators~\citep{lin1989robust}, may offer improved performance under heavy censoring, and represent a potential extension direction for future research.

It is worth mentioning that the TADA framework itself does not require assumptions such as proportional hazards to be valid in the target population. Rather, TADA’s objective is to reweight the study data so that the distribution of covariates aligns with that of the target population, thus facilitating an unbiased estimation of treatment effects at the target-population level. In the current implementation, we rely on the Cox PH model under the PH assumption, allowing us to interpret the TADA estimate as a constant hazard ratio. If non-proportional hazards are present, this Cox-based result may still be viewed as a time-averaged hazard ratio; yet, investigators seeking to capture time-varying effects more accurately could employ alternative outcome models, such as accelerated failure time models~\citep{collett2023modelling,kalbfleisch2002statistical}, piecewise Cox models~\citep{breslow1974covariance}, or additive hazards frameworks~\citep{lin1994semiparametric}. In such scenarios, the reweighting scheme of TADA remains applicable. Users simply need to replace the Cox PH with a more suitable outcome model, and TADA will yield a corresponding estimator that better reflects the underlying hazard structure in the target population.

Future research could expand TADA’s scope by investigating additional weighting strategies for achieving population balance and by exploring more flexible techniques for censoring adjustments. Although our current implementation uses the MoM to align study and target populations, and a standard form of IPCW to address censoring, there remains room to adopt censoring models that do not rely on proportional hazards or similarly restrictive assumptions~\citep{lee2024transporting}. Examining these alternatives could enhance TADA’s versatility in settings where the censoring patterns deviate from idealized conditions. Exploring approaches such as doubly robust or semiparametric methods would also help alleviate partial misspecification of either weighting component. Finally, while TADA has thus far been demonstrated using RCT data, extending it to accommodate other experimental and observational studies offers another avenue for broadening its applicability, particularly in cases where randomized data may be unavailable.

\section{Conclusion}
\label{sec9}
\setstretch{1.3}

Our study presents TADA’s performance to address censoring bias across various scenarios, demonstrating effective bias control and coverage even under challenging conditions. The TADA method’s design for using aggregate data for causal inference, along with its ability to accommodate varying censoring proportions, makes it a valuable tool for transporting survival estimates, especially in settings where target IPD is may not be available.

\subsection* {Acknowledgments}
The authors wish to thank Dr. Daniel Daly Grafstein for his contributions during the revision of the manuscript. The authors also thank the Editor and anonymous referees of this article for their valuable and thoughtful input. These suggestions have significantly enhanced the quality of the manuscript.

\subsection* {Conflict of Interest}
The authors declare no conflicts of interest.

\subsection* {Data Availability Statement}
All data used in the simulation study were generated according to the data-generating mechanisms specified in the manuscript. The individual-patient level data for the real case study are openly available in the Project Data Sphere dataset at https://doi.org/10.34949/05tc-jx78, reference number NCT00981058.

\newpage
\setstretch{1.25}
\bibliographystyle{WileyNJD-AMA}
\bibliography{main}

\begin{thebibliography}{10}
\providecommand \doibase [0]{http://dx.doi.org/}%

\bibitem{degtiar2023review}
Degtiar I, Rose S. A review of generalizability and transportability. {\it Annual Review of Statistics and Its Application} 2023\string; 10(1)\string: 501--524.

\bibitem{excellence2013guide}
Excellence C. Guide to the Methods of Technology Appraisal 2013 [Internet].  2013.

\bibitem{cole2010generalizing}
Cole SR, Stuart EA. Generalizing evidence from randomized clinical trials to target populations: the ACTG 320 trial. {\it American Journal of Epidemiology} 2010\string; 172(1)\string: 107--115.

\bibitem{stuart2011use}
Stuart EA, Cole SR, Bradshaw CP, Leaf PJ. The use of propensity scores to assess the generalizability of results from randomized trials. {\it Journal of the Royal Statistical Society Series A: Statistics in Society} 2011\string; 174(2)\string: 369--386.

\bibitem{atkins2011assessing}
Atkins D, Chang SM, Gartlehner G, et al. Assessing applicability when comparing medical interventions: AHRQ and the Effective Health Care Program. {\it Journal of Clinical Epidemiology} 2011\string; 64(11)\string: 1198--1207.

\bibitem{wang2019using}
Wang SV, Schneeweiss S, Gagne JJ, et al. Using real-world data to extrapolate evidence from randomized controlled trials. {\it Clinical Pharmacology \& Therapeutics} 2019\string; 105(5)\string: 1156--1163.

\bibitem{williams2020external}
Williams MJ. External validity and policy adaptation: From impact evaluation to policy design. {\it The World Bank Research Observer} 2020\string; 35(2)\string: 158--191.

\bibitem{buchanan2018generalizing}
Buchanan AL, Hudgens MG, Cole SR, et al. Generalizing evidence from randomized trials using inverse probability of sampling weights. {\it Journal of the Royal Statistical Society Series A: Statistics in Society} 2018\string; 181(4)\string: 1193--1209.

\bibitem{tipton2013stratified}
Tipton E. Stratified sampling using cluster analysis: A sample selection strategy for improved generalizations from experiments. {\it Evaluation Review} 2013\string; 37(2)\string: 109--139.

\bibitem{yang2020doubly}
Yang S, Kim JK, Song R. Doubly robust inference when combining probability and non-probability samples with high dimensional data. {\it Journal of the Royal Statistical Society Series B: Statistical Methodology} 2020\string; 82(2)\string: 445--465.

\bibitem{dahabreh2020extending}
Dahabreh IJ, Robertson SE, Steingrimsson JA, Stuart EA, Hernan MA. Extending inferences from a randomized trial to a new target population. {\it Statistics in Medicine} 2020\string; 39(14)\string: 1999--2014.

\bibitem{josey2021transporting}
Josey KP, Berkowitz SA, Ghosh D, Raghavan S. Transporting experimental results with entropy balancing. {\it Statistics in Medicine} 2021\string; 40(19)\string: 4310--4326.

\bibitem{josey2022calibration}
Josey KP, Yang F, Ghosh D, Raghavan S. A calibration approach to transportability and data-fusion with observational data. {\it Statistics in Medicine} 2022\string; 41(23)\string: 4511--4531.

\bibitem{quan2025generalizing}
Quan H, Li T, Chen X, Li G. Generalizing Treatment Effect to a Target Population Without Individual Patient Data in a Real-World Setting. {\it Pharmaceutical Statistics} 2025.

\bibitem{chen2023entropy}
Chen R, Chen G, Yu M. Entropy balancing for causal generalization with target sample summary information. {\it Biometrics} 2023\string; 79(4)\string: 3179--3190.

\bibitem{ramagopalan2022transportability}
Ramagopalan SV, Popat S, Gupta A, et al. Transportability of overall survival estimates from US to Canadian patients with advanced non--small cell lung cancer with implications for regulatory and health technology assessment. {\it JAMA Network Open} 2022\string; 5(11)\string: e2239874--e2239874.

\bibitem{zuo2022transportability}
Zuo S, Josey KP, Raghavan S, Yang F, Juar{\'e}z-Colunga E, Ghosh D. Transportability Methods for Time-to-Event Outcomes: Application in Adjuvant Colon Cancer Trials. {\it JCO Clinical Cancer Informatics} 2022\string; 6\string: e2200088.

\bibitem{lee2024transporting}
Lee D, Gao C, Ghosh S, Yang S. Transporting survival of an HIV clinical trial to the external target populations. {\it Journal of biopharmaceutical statistics} 2024\string; 34(6)\string: 922--943.

\bibitem{berkowitz2018generalizing}
Berkowitz SA, Sussman JB, Jonas DE, Basu S. Generalizing intensive blood pressure treatment to adults with diabetes mellitus. {\it Journal of the American College of Cardiology} 2018\string; 72(11)\string: 1214--1223.

\bibitem{lee2022doubly}
Lee D, Yang S, Wang X. Doubly robust estimators for generalizing treatment effects on survival outcomes from randomized controlled trials to a target population. {\it Journal of Causal Inference} 2022\string; 10(1)\string: 415--440.

\bibitem{cao2024transporting}
Cao Z, Cho Y, Li F. Transporting randomized trial results to estimate counterfactual survival functions in target populations. {\it Pharmaceutical Statistics} 2024\string; 23(4)\string: 442--465.

\bibitem{park2025introducing}
Park JJ, Vuong Q, Yan Y, Kittredge S, Metcalfe RK. Introducing TransportHealth and DAGDraw: User-Friendly Open-Source Software for Transportability and Generalizability Analyses and Causal Reasoning. {\it VeriXiv} 2025\string; 2\string: 67.

\bibitem{thatcher2015necitumumab}
Thatcher N, Hirsch FR, Luft AV, et al. Necitumumab plus gemcitabine and cisplatin versus gemcitabine and cisplatin alone as first-line therapy in patients with stage IV squamous non-small-cell lung cancer (SQUIRE): an open-label, randomised, controlled phase 3 trial. {\it The lancet oncology} 2015\string; 16(7)\string: 763--774.

\bibitem{seung2020real}
Seung S, Hurry M, Walton R, Evans W. Real-world treatment patterns and survival in stage IV non-small-cell lung cancer in Canada. {\it Current Oncology} 2020\string; 27(4)\string: e361.

\bibitem{rubin1974estimating}
Rubin DB. Estimating causal effects of treatments in randomized and nonrandomized studies. {\it Journal of Educational Psychology} 1974\string; 66(5)\string: 688--701.

\bibitem{rubin1986comment}
Rubin DB. Comment: Which ifs have causal answers. {\it Journal of the American Statistical Association} 1986\string; 81(396)\string: 961--962.

\bibitem{lee2023improving}
Lee D, Yang S, Dong L, Wang X, Zeng D, Cai J. Improving trial generalizability using observational studies. {\it Biometrics} 2023\string; 79(2)\string: 1213--1225.

\bibitem{hernan2010hazards}
Hern{\'a}n MA. The hazards of hazard ratios. {\it Epidemiology} 2010\string; 21(1)\string: 13--15.

\bibitem{ackerman2021generalizing}
Ackerman B, Lesko CR, Siddique J, Susukida R, Stuart EA. Generalizing randomized trial findings to a target population using complex survey population data. {\it Statistics in Medicine} 2021\string; 40(5)\string: 1101--1120.

\bibitem{signorovitch2010comparative}
Signorovitch JE, Wu EQ, Yu AP, et al. Comparative effectiveness without head-to-head trials: a method for matching-adjusted indirect comparisons applied to psoriasis treatment with adalimumab or etanercept. {\it Pharmacoeconomics} 2010\string; 28\string: 935--945.

\bibitem{phillippo2020equivalence}
Phillippo DM, Dias S, Ades A, Welton NJ. Equivalence of entropy balancing and the method of moments for matching-adjusted indirect comparison. {\it Research Synthesis Methods} 2020\string; 11(4)\string: 568--572.

\bibitem{hainmueller2012entropy}
Hainmueller J. Entropy balancing for causal effects: A multivariate reweighting method to produce balanced samples in observational studies. {\it Political Analysis} 2012\string; 20(1)\string: 25--46.

\bibitem{signorovitch2012matching}
Signorovitch JE, Sikirica V, Erder MH, et al. Matching-adjusted indirect comparisons: a new tool for timely comparative effectiveness research. {\it Value in Health} 2012\string; 15(6)\string: 940--947.

\bibitem{cox1972regression}
Cox DR. Regression models and life-tables. {\it Journal of the Royal Statistical Society Series B: Statistical Methodology} 1972\string; 34(2)\string: 187--202.

\bibitem{therneau2000cox}
Therneau TM, Grambsch PM, Therneau TM, Grambsch PM. {\it The cox model}.
\newblock Springer .
\newblock 2000.

\bibitem{breslow1972discussion}
Breslow NE. Discussion of Professor Cox's paper. {\it Journal of the Royal Statistical Society Series B: Statistical Methodology} 1972\string; 34\string: 216.

\bibitem{robins2000marginal}
Robins JM, Hernan MA, Brumback B. Marginal structural models and causal inference in epidemiology. {\it Epidemiology} 2000\string; 11(5)\string: 550--560.

\bibitem{hernan2000marginal}
Hern{\'a}n M{\'A}, Brumback B, Robins JM. Marginal structural models to estimate the causal effect of zidovudine on the survival of HIV-positive men. {\it Epidemiology} 2000\string; 11(5)\string: 561--570.

\bibitem{binder1992fitting}
Binder DA. Fitting Cox's proportional hazards models from survey data. {\it Biometrika} 1992\string; 79(1)\string: 139--147.

\bibitem{morris2019using}
Morris TP, White IR, Crowther MJ. Using simulation studies to evaluate statistical methods. {\it Statistics in Medicine} 2019\string; 38(11)\string: 2074--2102.

\bibitem{kalbfleisch2011statistical}
Kalbfleisch JD, Prentice RL. {\it The statistical analysis of failure time data}.
\newblock John Wiley \& Sons .
\newblock 2011.

\bibitem{collett2023modelling}
Collett D. {\it Modelling survival data in medical research}.
\newblock Chapman and Hall/CRC .
\newblock 2023.

\bibitem{green2015project}
Green AK, Reeder-Hayes KE, Corty RW, et al. The project data sphere initiative: accelerating cancer research by sharing data. {\it The oncologist} 2015\string; 20(5)\string: 464--e20.

\bibitem{lin1989robust}
Lin DY, Wei LJ. The robust inference for the Cox proportional hazards model. {\it Journal of the American Statistical Association} 1989\string; 84(408)\string: 1074--1078.

\bibitem{kalbfleisch2002statistical}
Kalbfleisch JD, Prentice RL. {\it The statistical analysis of failure time data}.
\newblock John Wiley \& Sons .
\newblock 2002.

\bibitem{breslow1974covariance}
Breslow N. Covariance analysis of censored survival data. {\it Biometrics} 1974\string: 89--99.

\bibitem{lin1994semiparametric}
Lin DY, Ying Z. Semiparametric analysis of the additive risk model. {\it Biometrika} 1994\string; 81(1)\string: 61--71.

\end{thebibliography}

\includepdf[pages=-]{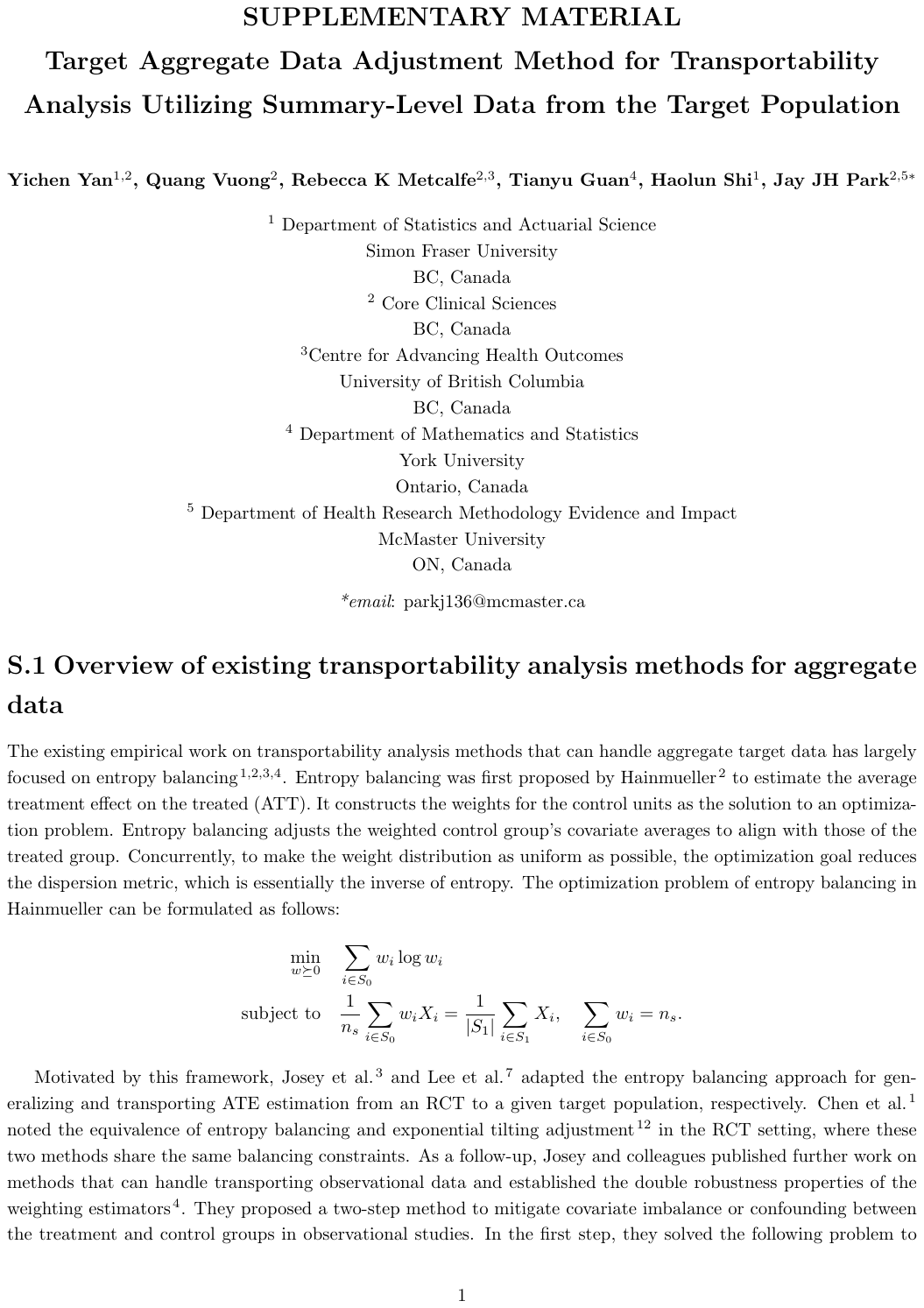}

\end{document}